\documentclass[aps,reprint,prb,twocolumn,showpacs,superscriptaddress]{revtex4-1}  
\usepackage{graphicx}  
\usepackage{dcolumn}   
\usepackage{bm}        
\usepackage{amssymb}   
\usepackage{amsmath}
\usepackage{gensymb}
\usepackage{color}
\usepackage{bbm}
\usepackage{hyperref}

\newcommand{\kapp}{(Cu$_{0.73}$Zn$_{0.27}$)$_3$(Zn$_{0.88}$Cu$_{0.12}$)(OH)$_6$Cl$_2$}

\begin{document}

\title{Spin dynamics and disorder effects in the $S=\frac{1}{2}$ kagome Heisenberg spin liquid phase of kapellasite}

\author{E.~Kermarrec}\affiliation{Laboratoire de Physique des Solides, Universit\'e Paris Sud, UMR CNRS 8502, F-91405 Orsay, France}
\author{A.~Zorko}\affiliation{Jo$\check{z}$ef Stefan Institute, Jamova 39, 1000 Ljubljana, Slovenia}
\affiliation{EN--FIST Centre of Excellence, Dunajska 156, SI-1000 Ljubljana, Slovenia}
\author{F. Bert}\affiliation{Laboratoire de Physique des Solides, Universit\'e Paris Sud, UMR CNRS 8502, F-91405 Orsay, France}
\author{R.H.~Colman}\affiliation{University College London, Department of Chemistry, 20 Gordon Street, London, WC1H 0AJ, UK}
\author{B.~Koteswararao}\affiliation{Laboratoire de Physique des Solides, Universit\'e Paris Sud, UMR CNRS 8502, F-91405 Orsay, France}
\author{F.~Bouquet}\affiliation{Laboratoire de Physique des Solides, Universit\'e Paris Sud, UMR CNRS 8502, F-91405 Orsay, France}
\author{P.~Bonville} \affiliation{Service de Physique de l'\'Etat Condens\'e, CEA-CNRS, CE-Saclay,  91191 Gif-Sur-Yvette, France}
\author{A.~Hillier} \affiliation{ISIS Facility, Rutherford Appleton Laboratory, Chilton, Didcot, Oxon OX11 OQX, UK}
\author{A.~Amato} \affiliation{Laboratory for Muon Spin Spectroscopy, Paul Scherrer Institut, CH-5232 Villigen PSI, Switzerland}
\author{J. van Tol}
\affiliation{National High Magnetic Field Laboratory, Florida State University, Tallahassee, Florida 32310, USA}
\author{A. Ozarowski}
\affiliation{National High Magnetic Field Laboratory, Florida State University, Tallahassee, Florida 32310, USA}
\author{A.S.~Wills}\affiliation{University College London, Department of Chemistry,  20 Gordon Street, London, WC1H 0AJ, UK}
\author{P.~Mendels}\affiliation{Laboratoire de Physique des Solides, Universit\'e Paris Sud, UMR CNRS 8502, F-91405 Orsay, France}
\affiliation{Institut Universitaire de France, 103 bd Saint-Michel, F-75005 Paris, France}
%

\begin{abstract}
We report $^{35}$Cl NMR, ESR, $\mu$SR and specific heat measurements on the $S=1/2$ frustrated kagom\'e magnet kapellasite, $\alpha-$Cu$_3$Zn(OH)$_6$Cl$_2$, where a gapless spin liquid phase is stabilized by a set of \textit{competing} exchange interactions.
Our measurements confirm the ferromagnetic character of the nearest-neighbour exchange interaction $J_1$ and give an energy scale for the competing interactions $|J| \sim 10$~K. The study of the temperature-dependent ESR lineshift reveals a moderate symmetric exchange anisotropy term $D$, with $|D/J|\sim 3$~\%. These findings validate \textit{a posteriori} the use of the $J_1 - J_2 - J_d$ Heisenberg model to describe the magnetic properties of kapellasite [Bernu \textit{et al.}, Phys. Rev. B \textbf{87}, 155107 (2013)].
We further confirm that the main deviation from this model is the severe random depletion of the magnetic kagom\'e lattice by 27\%, due to Cu/Zn site mixing, and specifically address the effect of this disorder by $^{35}$Cl NMR, performed on an oriented polycrystalline sample.
Surprisingly, while being very sensitive to local structural deformations, our NMR measurements demonstrate that the system remains homogeneous with a unique spin susceptibility at high temperature, despite a variety of magnetic environments.
Unconventional spin dynamics is further revealed by NMR and $\mu$SR in the low-$T$, correlated, spin liquid regime, where a broad distribution of spin-lattice relaxation times is observed. We ascribe this to the presence of local low energy modes.
\end{abstract}

\maketitle

\section{Introduction}
%
%
\begin{figure}
\includegraphics[width=\columnwidth]{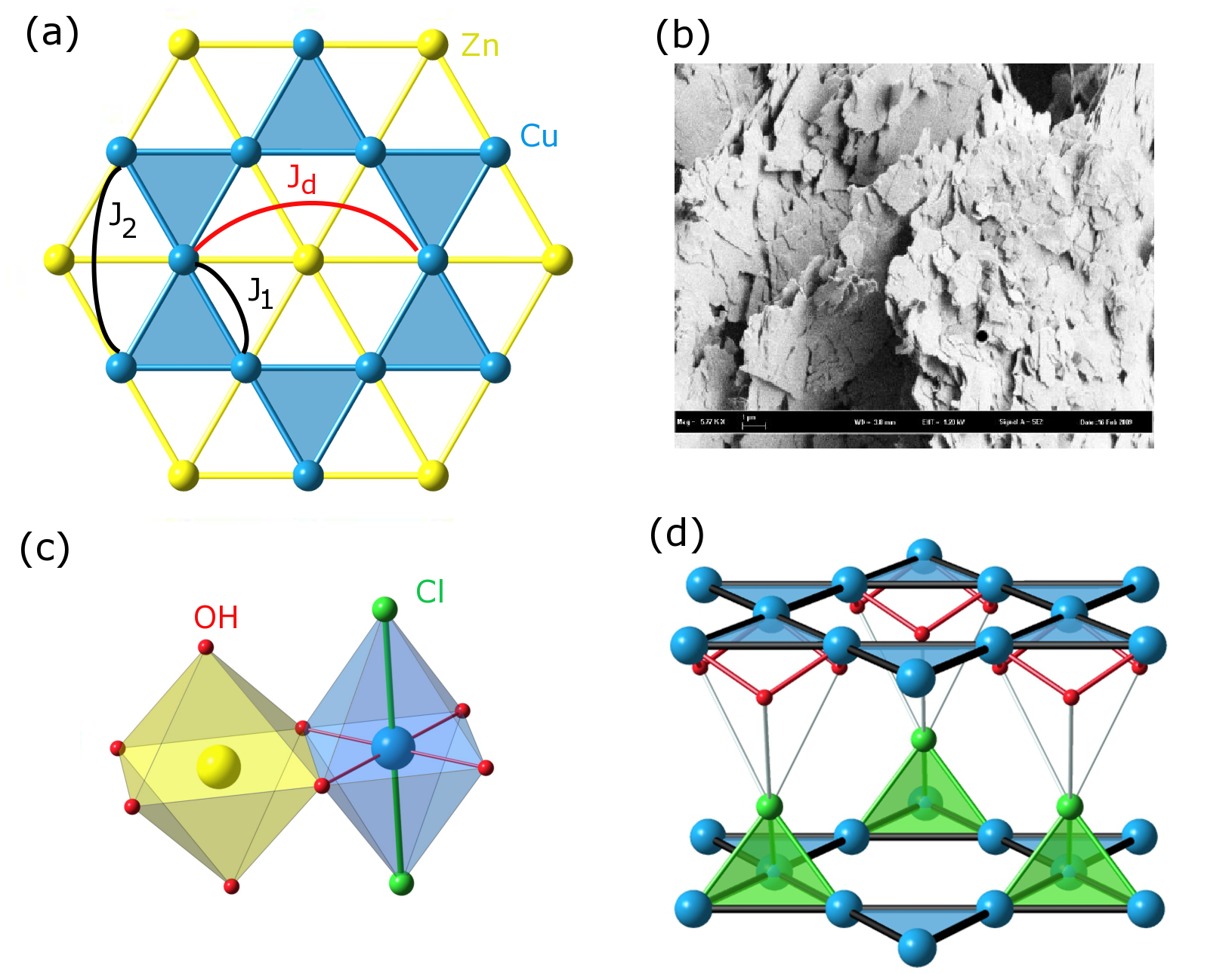}
\caption{\label{structure} (Color online) (a) The kagom\'e magnetic pattern arises from the regular arrangement of Cu$^{2+}$ ($S=1/2$) within a Cu/Zn triangular lattice (Cu in blue, Zn in yellow). The main exchange couplings $J_1$, $J_2$ and $J_d$ are presented on the diagram. (b) Scanning electron microscope image of kapellasite. The width of the image is 20~$\mu$m. The flat shape of the micron scale grains evidences the bidimensional character of kapellasite.
(c) Octahedral coordination environments of Cu and Zn.
(d) The kagom\'e layers are only weakly coupled through Cl-OH hydrogen bonds. The $^{35}$Cl NMR probe is dominantly coupled to three Cu. Zn atoms are not shown for clarity.}
\end{figure}
The $S=\frac{1}{2}$ Heisenberg antiferromagnet model on a kagom\'e lattice has proven to be a fascinating and enduring problem for more than 20 years. Currently, there seems to be a broad consensus on the nature of the ground state, with recent DMRG studies pointing to a resonating valence bond state that features a singlet-triplet gap, though the proposed value of $\Delta = 0.13J$\cite{Yan2011,Depenbrock2012} is still uncertain.\cite{Hotta2013} Contrastingly, fermionic approaches which suggest a gapless ground state remain debated.\cite{Iqbal2014,Hastings2000,Ran2007,Hermele2008}
%
Meanwhile, on the experimental side, new spin liquid candidates have been discovered in recent years with $S=\frac{1}{2}$ spins.\cite{Shores2005,Aidoudi2011,Colman2011,Kap_formula}  One of the most puzzling results, in view of the DMRG results, is the absence of any gap in the excitations of these experimental realizations. The reasons for such a discrepancy, and whether it is related to deviations from the ideal model, remain key open issues in the field. Several research directions have been proposed based on both theoretical developments and experiments such as Dzyaloshinskii-Moriya anisotropic interaction and next nearest-neighbours (nnn) exchange,\cite{Zorko,ColmanVes2011,RMN_Vesig,Zorko2013} which are believed to drive the system towards a magnetically ordered phase,\cite{Cepas2008,Messio2010,Huh2010} or lattice distortions.

In this landscape of new spin liquid candidates and drive to understand the role of perturbations to the ideal nearest-neighbour (nn) Heisenberg antiferromagnetic model, kapellasite,\cite{Kap_formula,Colman2008} a polymorph of herbertsmithite $\gamma-$Cu$_3$Zn(OH)$_6$Cl$_2$, has come as a surprise. Instead of being a mere variant of herbertsmithite, it has been demonstrated to be a kagom\'e spin liquid candidate of a new type. $\mu$SR and inelastic neutron scattering (INS) experiments evidenced a gapless spin liquid state down to 20~mK, with unusual dynamical short-range spin correlations. It is an experimental representative of a previously uninvestigated class of models on the kagom\'e lattice with frustrated \emph{competing} interactions, namely ferromagnetic nn-exchange $J_1$ and second nn (along the edges) $J_2$, and antiferromagnetic nnn exchange (across the diagonal of the hexagons) $J_d$, with $J_2\ll J_1 \sim J_d$.\cite{Kap_letter, Jeschke2013} The values of interactions extracted from a high-$T$ series analysis of macroscopic susceptibility and specific heat lead to an estimate of $J_1 = -12 $~K, $J_2 = -4 $~K and $J_d =15.6 $~K,\cite{Kap_letter,Bernu} which, for classical spins, would stabilize a non-coplanar chiral state of the cuboc2 type. In kapellasite, quantum fluctuations are argued to destabilize the latter state and to lead to related dynamical correlations, as detected by INS.\cite{Kap_letter}

The magnetic structure of kapellasite, $\alpha-$Cu$_3$Zn(OH)$_6$Cl$_2$, exhibits well decoupled magnetic planes with a kagom\'e geometry formed from the depletion of a Cu$^{2+}$ ($S=\frac{1}{2}$) triangular lattice by diamagnetic Zn$^{2+}$ [Fig.\ref{structure}(a)].
These planes are only weakly coupled along the third dimension (the hexagonal $c$-axis) via interlayer Cl$\cdots$OH hydrogen bonding, which guarantees highly bidimensional characteristics.
This essential feature causes the flat shape of micron scale grains constituting our powder sample [Fig.\ref{structure}(b)].
Neutron diffraction revealed a non-distorted kagom\'e lattice, but showed a sizeable shortage of Cu compared to the ideal stoichiometry.\cite{Kap_formula}
The observed formula determined by ICP--AES chemical analysis is \kapp.
Importantly, the local octahedral environment of nominally Cu site is more distorted than the Zn one, and has an extension along the Cl-Cl direction [Fig.\ref{structure}(c)].
The more symmetric bonding environment of the Zn site, an octahedral site coordinated by six oxygens, leads to strong preferential occupation by Zn$^{2+}$, with 95~\% of the Jahn--Teller active Cu$^{2+}$ residing on the more distorted kagom\'e sites coordinated by four oxygens and two chlorines.
The major consequence of this non-stoichiometry is then the depletion of the kagom\'e lattice by non-magnetic Zn$^{2+}$ ions. The kagom\'e site occupancy, $p=0.73$, still remains above the percolation threshold of the kagom\'e lattice $p_c \simeq 0.65$.\cite{Percolation1,Percolation2}

In this article, we investigate the magnetic properties and the effects of structural disorder in kapellasite by means of NMR, $\mu$SR, ESR and specific heat measurements.
This paper is organized as follows.
In section \ref{NMR}, we evaluate by $^{35}$Cl-NMR the effect of disorder within the kagom\'e planes and report evidence for local homogeneity of the magnetic properties, proving the robustness of the underlying physics against dilution.
In section \ref{ESR}, we use ESR to estimate the symmetric exchange anisotropy term in the Hamiltonian and confirm that the choice of an isotropic Heisenberg Hamiltonian to describe the magnetic properties of kapellasite is  reasonable in a first approach.
While the main exchange interactions $J_1$ and $J_d$ have already been derived from magnetic susceptibility and specific heat fits, we detail in section \ref{Thermo} an analysis of the specific heat under magnetic fields which confirms that its low-$T$ part is magnetic
and intrinsic, and thus provides a pertinent quantity for extracting the exchange integrals.
%
Finally, in the last section \ref{Dynamics}, we investigate with NMR and $\mu$SR the spin fluctuations at low energy in the low-$T$ frustrated antiferromagnetic phase induced by these competing interactions on the $S=\frac{1}{2}$ kagom\'e lattice.

\section{\label{NMR}NMR determination of local structure and spin susceptibility}
NMR experiments were performed on the most abundant Cl isotope, $^{35}$Cl (abundance of 75.8~\%) using an oriented powder aligned along the hexagonal $c$-axis of the kapellasite's rhombohedral structure. The high quality orientation leads to accurate refinements of the local structure and unambiguous determination of the local spin susceptibility and the spin dynamics.

$^{35}$Cl is a quadrupolar nucleus ($I=3/2$), hence one has to consider two sets of interactions resulting from the coupling of the nucleus to its magnetic and charge environments. The nuclear Hamiltonian can be written\cite{slichter,cohen}
\begin{equation}
\mathcal{H} = \mathcal{H}_\mathrm{mag} + \mathcal{H}_\mathrm{quad} \\
\end{equation}
with $$ \mathcal{H}_\mathrm{quad} \gg \mathcal{H}_\mathrm{mag}  $$
As detailed in Appendix \ref{NMRappendix}, this generally leads for a given environment to a spectrum consisting of three lines. The central line, corresponding to the $-\frac{1}{2}\leftrightarrow\frac{1}{2}$ transition, and the satellite lines, corresponding to $-\frac{3}{2}\leftrightarrow -\frac{1}{2}$ and $\frac{1}{2}\leftrightarrow\frac{3}{2}$ transitions, which are sensitive to quadrupolar perturbations in second and first order respectively. More details about this nuclear Hamiltonian can be found in Appendix \ref{NMRappendix}.

\begin{figure}
\includegraphics[width=\columnwidth]{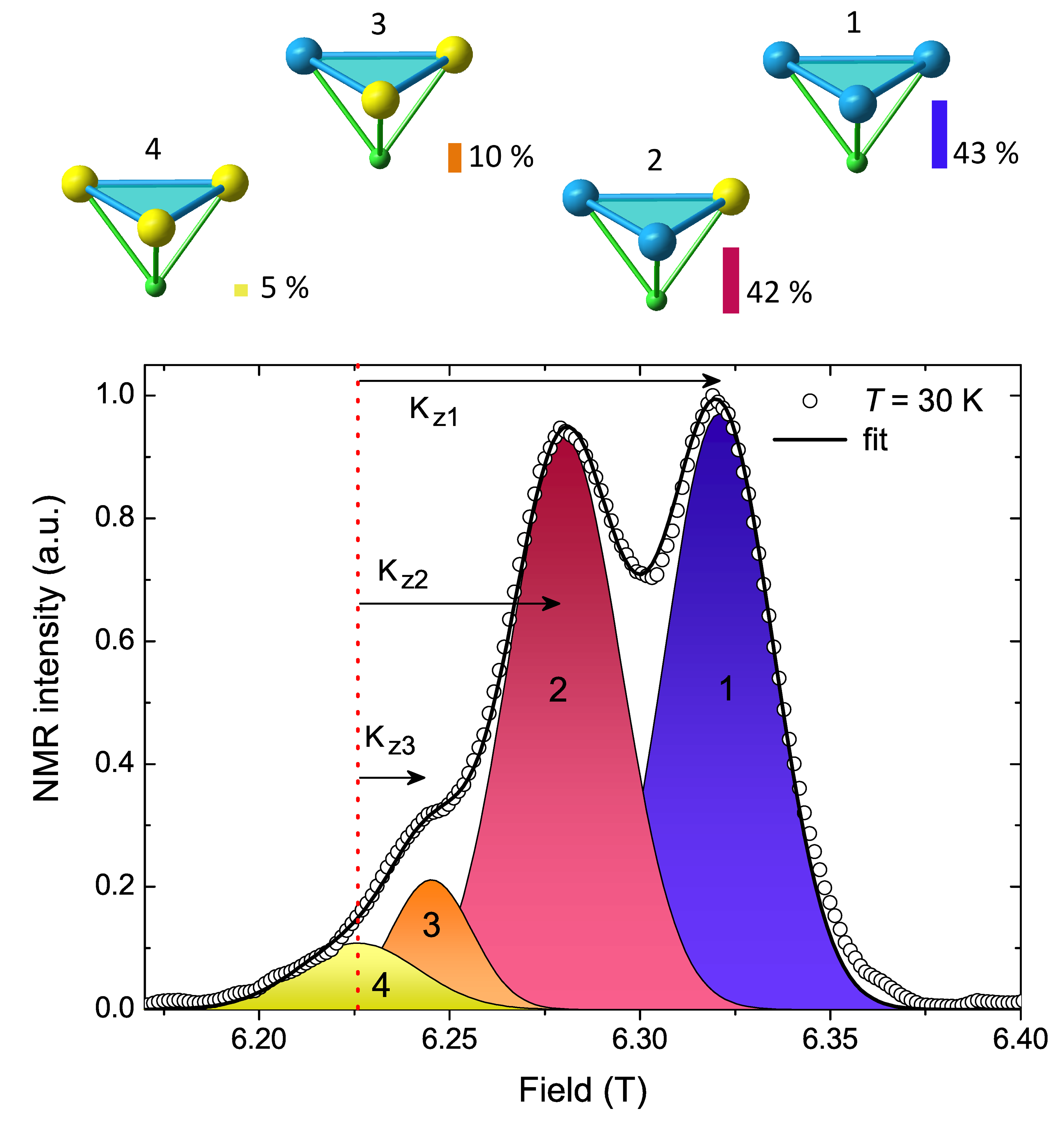}
\caption{\label{NMRlocal} (Color online) $^{35}$Cl NMR spectrum of the $-\frac{1}{2} \leftrightarrow \frac{1}{2}$ transition lines at $T = 30$~K, $\theta=0$\degree (empty circles). The shift reference is indicated by the red vertical dotted line. The NMR spectrum is fitted using four gaussian lines (black line), corresponding to various  $^{35}$Cl environments due to Cu/Zn (blue/yellow) mixing. The integrated intensity  $A_i$ of each line $i$ reveals the site probability.}
\end{figure}

\subsection{\label{Mat}Local structure}
\begin{figure*}
\includegraphics[width=2\columnwidth]{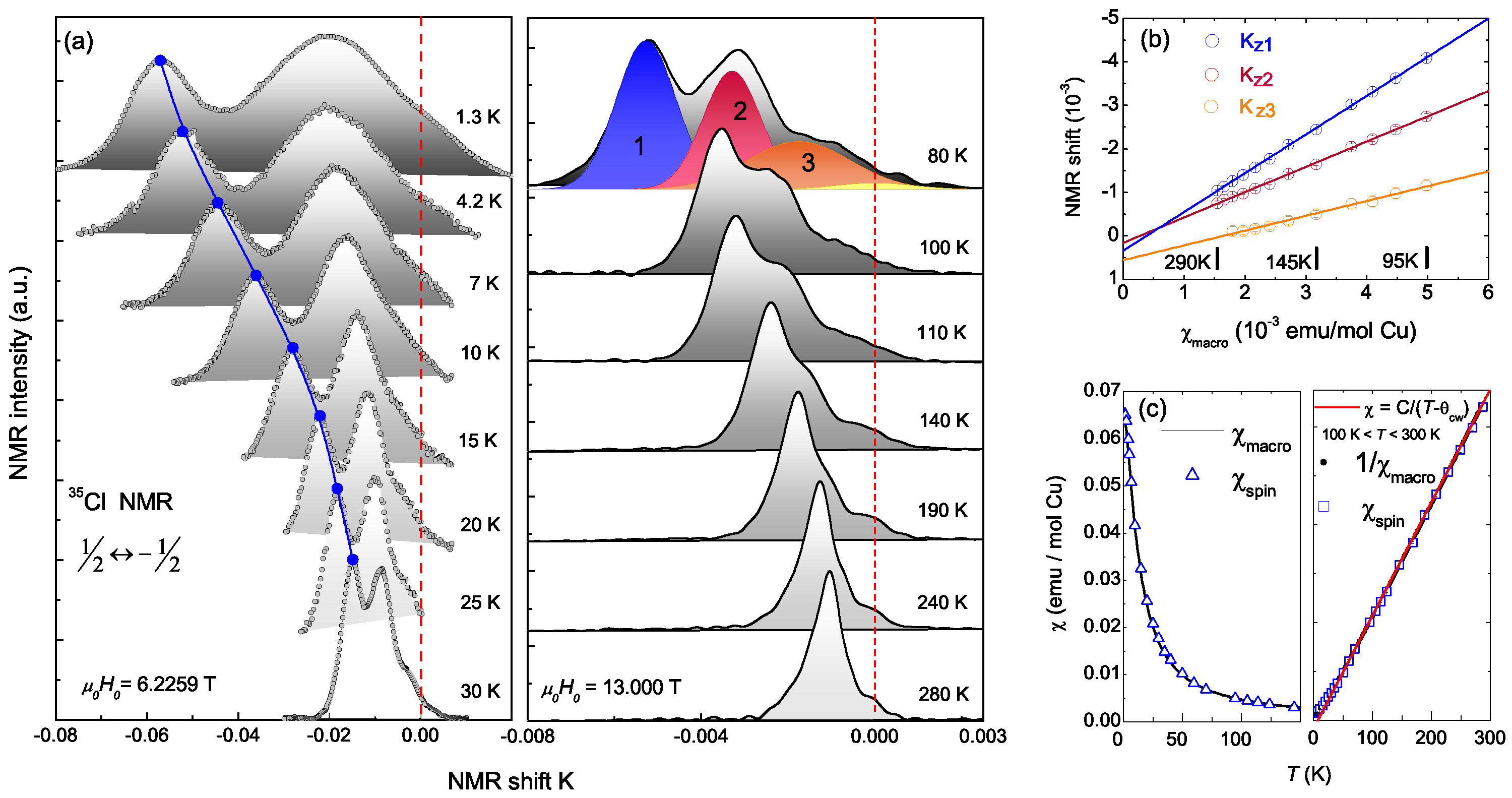}
\caption{\label{NMRchi} (Color online) (a)$^{35}$Cl NMR spectra of the central lines plotted versus the lineshift from high (right) to low (left) temperature. The spectra are vertically shifted for clarity. The composition of the spectrum in terms of different Cl sites is shown at 80~K. The local susceptibilites can be directly extracted from the shift of the lines from the reference (dashed vertical red lines). As an example, the blue circles emphasize the shift of the NMR line corresponding to site (1) (left panel).
(b) $^{35}$Cl NMR shifts for the different Cl sites vs. $\chi_\mathrm{macro}$ at high $T$. The hyperfine couplings are extracted from the slope of the curves ($A_\mathrm{hf,1}=-4.98(3)$~kOe/$\mu_{\rm B}$).
(c)Temperature dependence of the macroscopic susceptibility $\chi_\mathrm{macro}$, the local NMR susceptibility $\chi_\mathrm{spin}$ (left) and of $1/\chi_\mathrm{macro}$, $1/\chi_\mathrm{spin}$ (right). The Curie-Weiss parameter $\theta_\mathrm{CW} = 9.5\pm 1$~K evidences a weak ferromagnetic effective coupling at high $T$. }
\end{figure*}

In this section we discuss the local structure of kapellasite as deduced from $^{35}$Cl NMR experiments.
For a perfect coverage of the kagom\'e lattice with Cu$^{2+}$ ions, one would expect a unique Cl environment, where each Cl nucleus couples to the three adjacent Cu sites located at the corners of a given triangle. In disagreement with this simple scheme, we instead observe multiple lines, which are best resolved for the  {\it central} $-\frac{1}{2}\leftrightarrow\frac{1}{2}$ transition line at $T=30$~K taken for $\theta=0$\degree, where $\theta$ is the angle between the applied field and the $c$-axis (see Fig.\ref{NMRlocal} and section \ref{NMRsusceptibility}).
The multiple lines can be associated with different Cl environments induced by the Cu/Zn mixing disorder, namely four different Cu/Zn configurations on a probed triangle: Cu$_3$ (1),  Cu$_2$Zn (2), CuZn$_2$ (3) and Zn$_3$ (4). This site assignment also agrees with the high-$T$ shifts which scale with the number of Cu to which each Cl is coupled, see Fig.\ref{NMRchi}(b).
Since the spin-spin ($T_2$) and spin-lattice ($T_1$) corrections were found to be negligible, the integrated intensity of each line is directly proportional to the number of triangles of each configuration which is found to perfectly match with a random occupation of the Cu sites (Table \ref{Occupancies}), with a filling ratio $p=0.75(1)$. This agrees with the value of 0.73 extracted from neutron diffraction refinements~\cite{Kap_formula}.
\begin{figure}
\includegraphics[width=0.9\columnwidth]{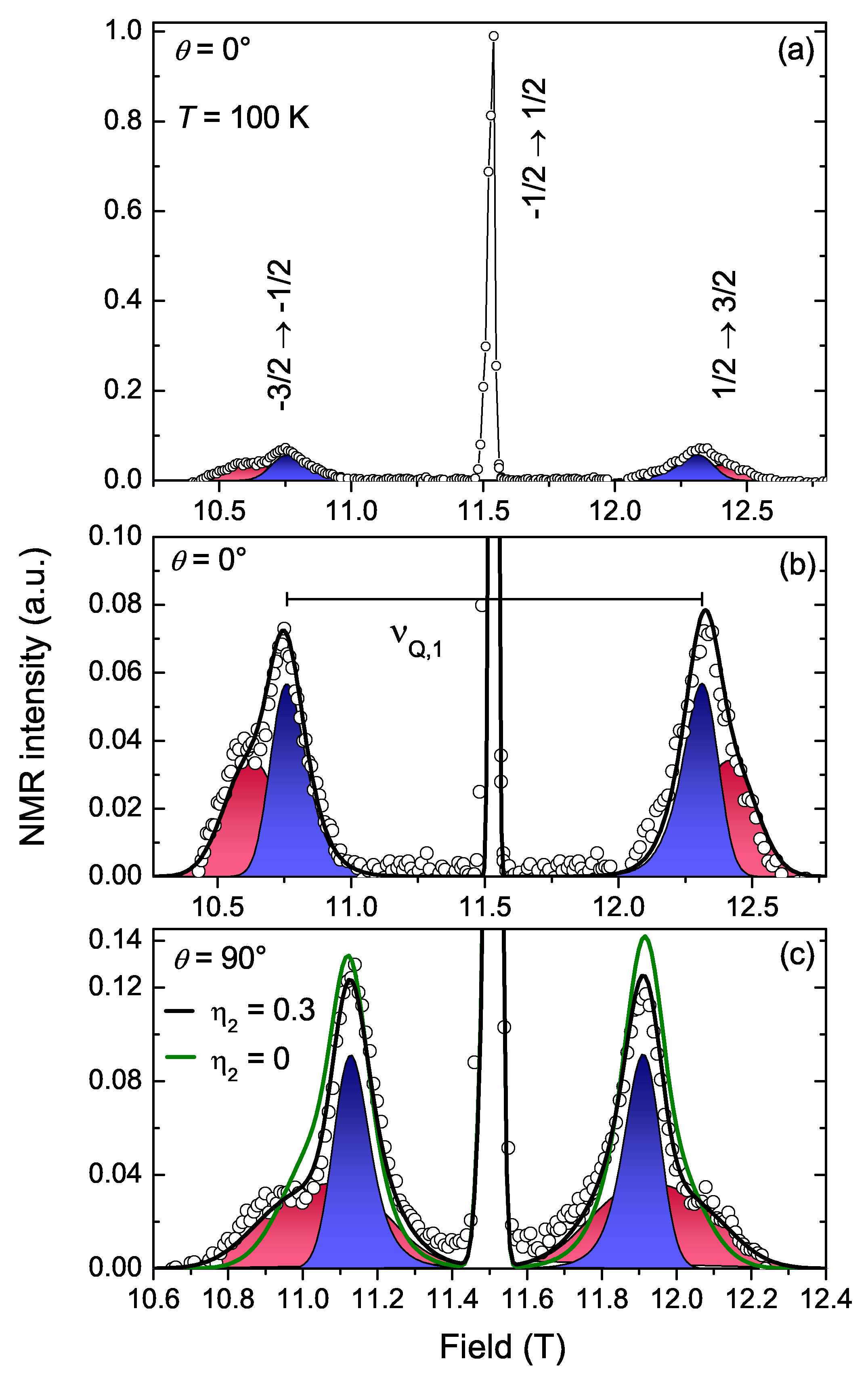}
\caption{\label{NMRfull} (Color online) (a)$^{35}$Cl NMR spectrum at $\nu_0 = 47.950$~MHz and $T = 100$~K. Three distinct groups of NMR lines are observed. (b) and (c): NMR satellites in two different angle configurations $\theta = 0$\degree (b) and $\theta = 90$\degree (c) at $T=100$~K. The total fit (black line) is the sum of the two satellites pairs of the main sites (site 1 in blue, site 2 in red). For $\theta = 90$\degree, the simulation with $\eta_2 = 0$ (green line) fails to represent the satellites of the site 2, contrary to the case where $\eta_2 = 0.3(1)$ (black line).}
\end{figure}
\begin{table}
\caption{\label{Occupancies} Cl sites occupancies evaluated by NMR lines integrals $A_i$ and compared to binomial probabilities assuming a random disorder  for a kagom\'e site filling ratio $p = 0.75$. }
\begin{ruledtabular}
\begin{tabular}{lccc}
Sites $i$ / config.          & Int. $A_i$          & Bin. prob. ($p=0.75$)          \\
\hline
\hline
1  ~Cu$_3$           & 0.43(1) & 0.42 & \\
2  ~Cu$_2$Zn      & 0.42(1) & 0.42 & \\
3  ~CuZn$_2$      & 0.10(5) & 0.14 & \\
4   ~Zn$_3$           & 0.05(4) & 0.02 & \\
\end{tabular}
\end{ruledtabular}
\end{table}

The impact of dilution on the local structure can be traced by the {\it satellite} lines which positions depend to first order upon the parameters which govern $\mathcal{H}_\mathrm{quad}$,  namely the quadrupolar frequency $\nu_q$ and the asymmetry $\eta$, and the orientation of the principal axis of the electric field gradient (EFG) - see Appendix \ref{NMRappendix}. Hence they reflect the difference in electrostatic environments at the local scale.
A complete NMR spectrum measured in the $\theta=0$\degree orientation is shown in Fig.\ref{NMRfull}(a).
Three groups of NMR lines are clearly observed which correspond to the three expected transitions for a spin $I=3/2$.
Only the satellite contribution from the two dominant main sites (1) and (2) can be reliably followed since the sites (3) and (4) represent a minor fraction (altogether 16~\%) of the total spectrum intensity.
Further details on satellites (1) and (2) are displayed in Fig.\ref{NMRfull}(b) and (c).
For $\theta=0$\degree, if one assumes that the principal axis of the EFG remains along the symmetry axis $Oz$ for all the Cl sites, the distance between the two satellites should be $2\nu_q$ (eqs.\ref{nu2},\ref{nu3}), independently of the $\eta$  value.
Under the latter assumption, we find two different values for sites (1) and (2), respectively $\nu_{q,1} = 3.4(1)$~MHz and $\nu_{q,2} = 4.0(2)$~MHz [Fig.\ref{NMRfull}(b)], which reflect two different local electrostatic environments around Cl. While for site (1) the spectrum is consistent with a $\eta = 0$ value, for site (2) the asymmetry of the spectrum obtained in the perpendicular direction ($\theta =90$\degree) can only be accounted for by assuming a surprisingly large value of the asymmetry parameter $\eta = 0.3(1)$. Finding such a large value might rather invalidate our initial assumption and indicate that for this site, the EFG and shift tensor principal axis do not coincide anymore.
The local distortions deduced through this extensive analysis of the quadrupolar parameters can easily be attributed to the different nature of the two Cu$^{2+}$ and Zn$^{2+}$ ions. Whereas they possess the same charge and a similar radius, Cu$^{2+}$ is known to be Jahn-Teller active and to elongate the octahedron built on the nearby anions. Such local effects can then be expected when the ratio Zn/Cu is varied on a given triangle but do not affect the average threefold axis structural symmetry.

\subsection{\label{NMRsusceptibility} Spin susceptibility}
Below we extend the concise presentation of Ref.\onlinecite{Kap_letter} and detail the analysis of the $^{35}$Cl NMR data taken on the central line, and further compare the susceptibilities extracted for sites (1), (2) and (3) from the lineshifts taken for an applied field along the $c$-axis,  $K_{z,1}$, $K_{z,2}$ from 287 K down to 1.2~K and $K_{z,3}$ down to 80~K (Fig.\ref{NMRchi}).
The broadening of the NMR lines observed at low temperature is discussed in section \ref{disorder}.

Following a typical two-step procedure, we derive the NMR spin shifts which are  induced by the local susceptibilities denoted $\chi_{i}$: from a linear high-$T$ fit of $K_{z,i}$ plotted versus the macroscopic susceptibility $\chi_\mathrm{macro}$ measured by SQUID magnetometry, $K_{z,i}=A_\mathrm{hf,i} \cdot \chi_\mathrm{macro} + \sigma_{i}$,
we extract the hyperfine coupling constants $A_\mathrm{hf,i}$ and the chemical shifts $\sigma_\mathrm{i}$ [Fig.\ref{NMRchi}(b)].
The local susceptibilities are then given by $\chi_{i}=(K_{z,i}-\sigma_i)/A_\mathrm{hf,i}$. The ratios of hyperfine couplings $A_\mathrm{hf,1}/A_\mathrm{hf,2} = 1.53(2)$ and $A_\mathrm{hf,3}/A_\mathrm{hf,2} = 0.58(2)$, with $A_\mathrm{hf,1}=-4.98(3)$~kOe/$\mu_{\rm B}$, are found to be in good agreement with the ratios of the number of Cu$^{2+}$ probed, respectively 1.5 and 0.5.
This further strengthens the previous analysis of the local structure, and again indicates that at high $T$ the $S=1/2$ spins are rather uncorrelated, depicting a fluctuating paramagnetic state. This validates the overall consistency of our analysis.

The local susceptibility derived for site (1) is certainly the least perturbed by defects and is the closest to that of the ideal system. It is found to scale with $\chi_\mathrm{macro}$ over the whole temperature range, down to 1.2~K, which invalidates the possibility of a dominant contribution of defects at low-$T$ in $\chi_\mathrm{macro}$ [Fig.\ref{NMRchi}(b)]. One should note that this finding strongly contrasts with the case of many correlated oxides where a Curie-like tail in macroscopic susceptibility is commonly attributed to defects, either spin vacancies in the magnetic lattice or other species of weakly coupled spins. In these systems, the disorder contributes to the  NMR linewidth only, and not to the lineshift.\cite{RMN_Olariu,RMN_Vesig,Limot1,Bono1}
This uncommon divergence as $T\rightarrow 0$ of $\chi_\mathrm{spin}$ in comparison with the reported $\chi_\mathrm{spin}$ of nn-Heisenberg kagom\'e antiferromagnets~\cite{RMN_Olariu,RMN_Vesig,RMN_volb} could only be explained within the framework of competing interactions on the kagom\'e lattice, $J_1 = -12 $~K, $J_2 = -4 $~K and $J_d =15.6 $~K, evaluated by high temperature series expansion as emphasized in Ref.\onlinecite{Kap_letter, Bernu}.
%
%
The Curie-Weiss temperature $\theta_{\rm CW} \sim 9 $~K is consequently the result of an averaging of these competing interactions.

\begin{figure}
\includegraphics[width=\columnwidth]{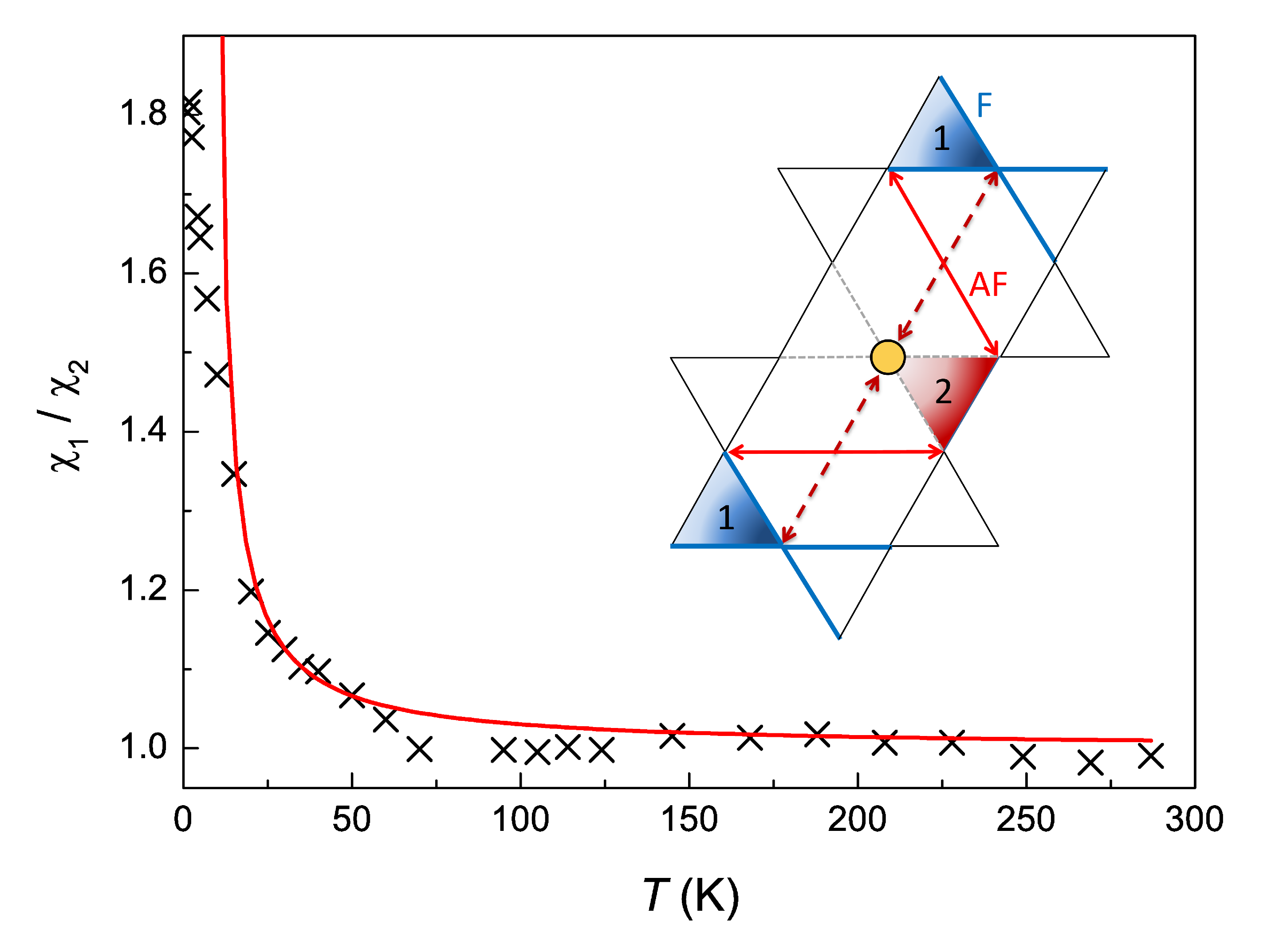}
\caption{\label{ratio} (Color online) Ratio of local susceptibilities $\chi_{\rm 1}/\chi_{\rm 2}$ (black crosses) vs. $T$. The red line is a fit using the ratio $(T - \theta_2) / (T - \theta_1)$ for $20 \leq T \leq 70$~K, with $\theta_1 = 7.8$~K and $\theta_2 = 5$~K (see text). The expected value of 1 is found above 70~K while the steep increase at low $T$ underlines the strengthening of antiferromagnetic correlations at the diluted site that is expected when $J_1$ is ferromagnetic.}
\end{figure}
A closer comparison of the low $T$ local susceptibilities $\chi_{i}$ at sites (1) and (2) brings an additional insight into the nature of the interactions. While all shifts display a low-$T$ upturn, a minor effect is detected below $T = 50$~K, where the ratio of the local susceptibilities of site (1) and site (2), $\chi_{\rm 1}/\chi_{\rm 2}$, is not $T$-independent anymore and shows instead a clear increase as $T$ is lowered (Fig.\ref{ratio}).
In order to infer a physical understanding to this deviation, $\chi_{\rm 1}/\chi_{\rm 2}$ is fitted to the ratio of two Curie-Weiss susceptibilities $C (T - \theta_2) / C (T - \theta_1)$.
The parameters $\theta_1$ and $\theta_2$ account for the interactions respectively probed at sites (1) and (2) and are expected to be relevant in the mean field regime, i.e. for $T \geqslant 2 \theta_{\rm CW} \simeq 20$~K.
The main result of this simple analysis is that the upturn at low $T$ could only be reproduced when $\theta_1 > \theta_2$.
In a mean field approach and ignoring the weakest $J_2$ coupling, the Curie-Weiss temperature of each site (\textit{i}) of the $J_1- J_d$ Heisenberg model on a diluted kagom\'e lattice can be simply expressed as $\theta_{i} = -\alpha_i J_1 - \beta_i J_d/2$. For a fully occupied kagom\'e lattice, one has $\alpha_{1,2} = \beta_{1,2} = 1$. Taking into account the random depletion of the lattice by 27~\%, a straightforward calculation of probabilities leads to $\alpha_{1} =0.865$,  $\beta_{1} = 0.73$ and $\alpha_{2} =0.615$,  $\beta_{2} = 0.73$. Therefore, one can directly infer from the relation $\theta_1 > \theta_2$ that $J_1 <0$ (ferromagnetic), in agreement with Ref. \onlinecite{Kap_letter, Bernu}.

The related scheme of interactions is summarized in the diagram of Fig.\ref{ratio}.

\section{\label{ESR}Electron Spin Resonance}
ESR measurements were performed at NHMFL, Tallahassee, Florida, on a custom-made ESR spectrometer with homodyne detection working in transmission mode at the fixed frequency of $\nu_0=326.4$~GHz. The same oriented sample was used as in NMR. A single ESR line was observed in the temperature range between 5 and 100~K. We notice first that this is consistent with the absence of any magnetic parasitic phase in the sample. Moreover, since the copper ions are present on both the nominally Cu and Zn sites, the observation of a single line indicates that there is a sizable exchange coupling between sites, yielding a Lorentzian line shape typical for an exchange narrowing limit.

There are three distinct $T$-regimes as revealed in (Fig.\ref{figESR}). Above 100~K, the ESR line could not be followed reliably due to its decreasing intensity with increasing temperature. The reduction of the signal amplitude was additionally enhanced by line broadening. Between 10 and 100~K (intermediate $T$), the linewidth increases regularly with increasing $T$. Below 10 K, an opposite behaviour is observed. In the two latter regimes, the resonance field exhibits a monotonic variation, similar to the macroscopic susceptibility.

Although the ESR line in kapellasite is similarly broad to that seen in herbertsmithite which points to a sizable anisotropy,\cite{Zorko} the temperature dependence of the ESR parameters; i.e., the line position $B_0$ ($g$-factor; $g=\frac{\nu_0}{\mu_B B_0}$, where $\mu_B$ is the Bohr magneton) and the linewidth $\Delta B$, are quite different. In herbertsmithite, a maximum in $\Delta B$ and small temperature changes of the $g$-factor of less than 0.02 were observed. Contrastingly, in kapellasite, a minimum is observed in the linewidth at 10~K. This behaviour rather resembles the one found in another Cu-based kagom\'e compound, vesignieite, where it was assigned to a phononic contribution.\cite{Zorko2013} Lastly, we note that the measured line-shifts with decreasing temperature in kapellasite are much larger than in herbertsmithite and vesignieite. This indicates that a different type of magnetic anisotropy is responsible for these shifts than the Dzyaloshinsky-Moriya (DM) interaction that had been found to be dominant in the other two cases. (Fig.\ref{figESR}).
Below, we discuss successively the information which can be extracted from the ESR linewidth and from the $g$-factor.

\subsection{Linewidth}
\begin{figure}[b]
\includegraphics[trim = 3mm 50mm 0mm 25mm, clip, width=1\linewidth]{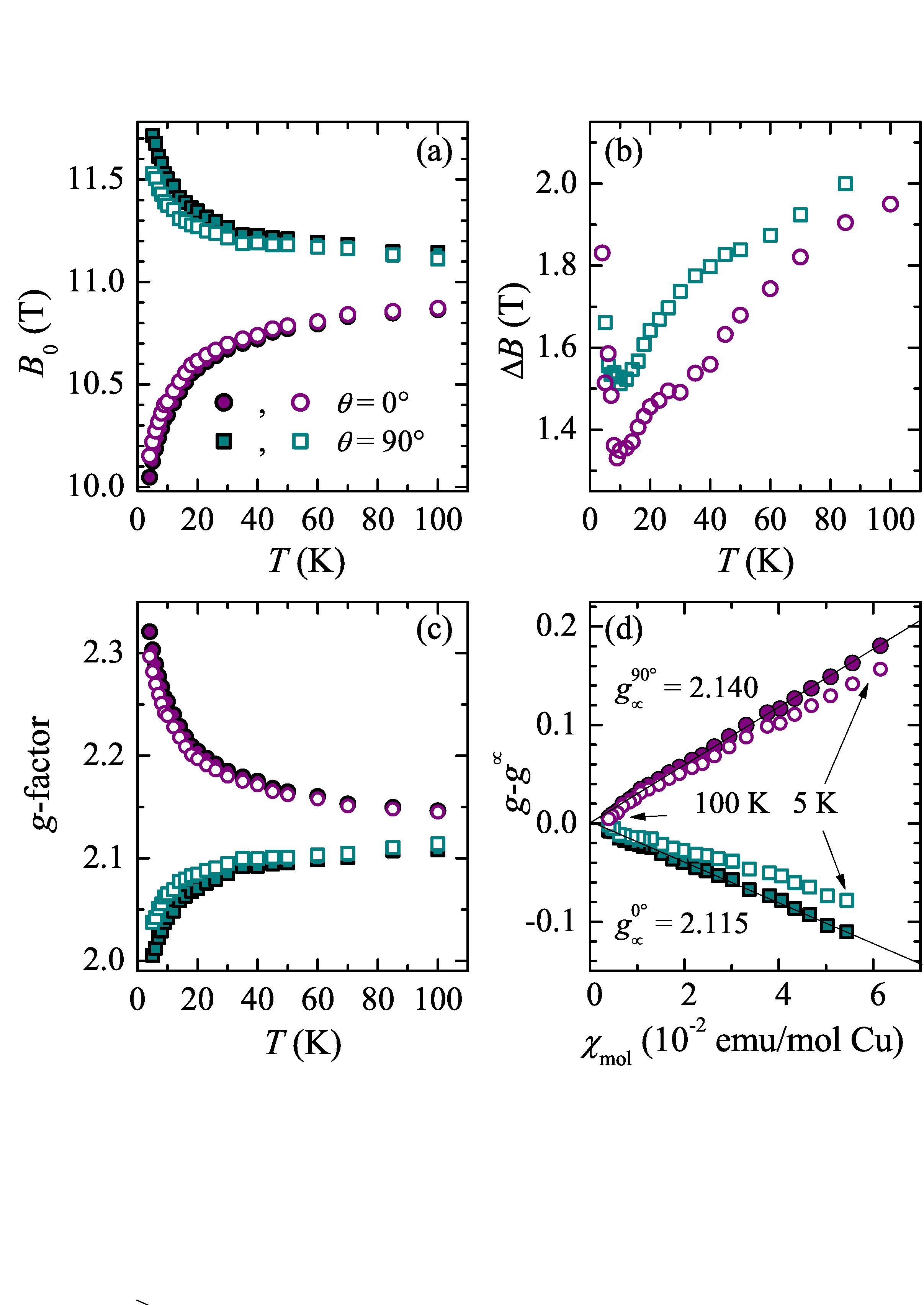}
\caption{Temperature dependence of (a) the ESR line position (b) linewidth and (c) $g$-factor measured at 326.4 GHz. (d) scaling of the g-factor with bulk susceptibility. Open symbols include all magnetic anisotropy contributions present in kapellasite, while solid symbols represent data corrected for the dipolar (including demagnetization) fields.}
\label{figESR}
\end{figure}

In the paramagnetic regime of herbertsmithite ($T\gg J$), the ESR linewidth was constant so this asymptotic value could be used to extract the dominant magnetic anisotropy term.\cite{Zorko} It is notable that, in kapellasite, one cannot reach this regime even at 100~K, a temperature already as high as $\sim 6J_d$ [Fig.\ref{figESR}(b)].  This indicates that the linewidth in the intermediate-$T$ regime cannot be accounted for solely by the spin system and an additional line-broadening mechanism must originate from other degrees of freedom. A spin-phonon coupling, responsible for a similar increase in vesignieite,\cite{Zorko2013} is quite commonly encountered in copper-based systems and is perfectly in line with the {\it increase} of the linewidth between 10 and 100~K.\cite{Seehra,Nidda, ZorkoSCBO, Herak}. This mechanism is dominant at high-$T$ so that the magnetic anisotropy cannot be extracted from the linewidth. The low-$T$ regime (below 10~K $\sim J$) resembles more the behavior found in herbertsmithite. At such low temperatures the phonon population vanishes and the ESR broadening becomes dominantly spin-induced, reflecting the development of spin correlations.

\subsection{$g$-factor}
Large changes of the $g$-factor with decreasing temperature in kapellasite [Fig.\ref{figESR}(c)] could, in principle, be due to either a locally inhomogeneous susceptibility or a significant magnetic anisotropy of the system. The first possibility can be easily discarded. Indeed, in inhomogeneous spin systems (e.g., spin glasses, impure systems and systems with several non-equivalent sites) nonuniform local fields develop with increasing susceptibility (decreasing $T$) leading exclusively to negative $B_0$ shifts, irrespective of the orientation of the applied field.\cite{Malozemoff,NagataJMMM} In addition line-broadening that scales with the susceptibility is generally encountered. Such shifts and broadening result from staggered local fields.\cite{NagataJMMM}
In kapellasite this is clearly not the case since both positive and negative line shifts are observed depending on the orientation of the applied field and the ESR width is non-monotonic when the susceptibility is increasing steeply at low temperature. We are then led to assign these shifts to magnetic anisotropy.

When the magnetic anisotropy $\mathcal{H}'$ is small compared to the exchange and/or Zeeman energy Nagata's theory for $g$-factor shifts can be applied.\cite{Nagata} Then, the shift is determined by the correlation function \cite{Nagata,Maeda}
\begin{equation}
\Delta g = \frac{\left\langle \left[ S^-,\left[S^+,\mathcal{H}' \right] \right] \right\rangle}{2\mu_B B_0\left\langle S^z\right\rangle},
\label{shift}
\end{equation}
where $\left[...,...\right]$ denotes a commutator, $\langle ... \rangle$ canonical averaging and $S^\alpha=\sum_i S_i^\alpha$ the $\alpha$-component of the total spin operator. Here, we refer to the general form of the magnetic anisotropy in the spin Hamiltonian of a spin-1/2 system,
\begin{equation}
\mathcal{H}'=\frac{1}{2}\sum_{i,j\neq i}{\bf S}_i\cdot {\underline K}_{ij} \cdot {\bf S}_j,
\label{anisoH}
\end{equation}
where the coupling tensor
\begin{equation}
{\underline K}_{ij}={\underline d}_{ij}+{\underline D}_{ij}
\label{aniso}
\end{equation}
includes both the long-range dipolar coupling ${\underline d}_{ij}$ and the short-range symmetric anisotropic exchange ${\underline D}_{ij}$ on bonds $J_1$ and $J_d$. We note that the antisymmetric Dzyaloshinsky-Moriya interaction does not contribute to the ESR line shift at the level of first order [eq.(\ref{shift})]\cite{Maeda} and so we omit it.

We proceed to the evaluation of both anisotropy terms from eq.(\ref{aniso}) in kapellasite, by noting that in the paramagnetic regime the Nagata's theory predicts a linear scaling of the line shift with susceptibility,\cite{Nagata,Nagata2}
\begin{equation}
\Delta g^z(T) = \frac{\langle S^z\rangle}{2\mu_B B_{0}}\sum_{j\neq i}{\left(2 K_{ij}^{zz}-K_{ij}^{xx}-K_{ij}^{yy}\right)},
\label{shift2}
\end{equation}
where $\langle S^z\rangle=\sum_i\langle S_i^z\rangle=\frac{\chi_{\rm mol}(T)B_0}{N_A g \mu_0\mu_B}$, $z$ denotes the orientation of the applied field and $x,y$ the two perpendicular directions, $N_A$ is the Avogadro number and $\mu_0$ the vacuum permeability. A convincing scaling is observed in kapellasite for both directions in the whole investigated temperature range [Fig.\ref{figESR}(d)], an additional proof that the increase of the bulk susceptibility is intrinsic even at low temperatures.
The scaling yields the limiting infinite-temperature values of the $g$-factor within the kagom\'e planes, $g_\infty^{90^\circ} =2.140 $, and along the $c$ axis, $g_\infty^{0^\circ}  =2.115 $. These values correspond to the values averaged over the structural triangular unit.

The contribution to the above sum related to the dipolar field arising from the finite spin $\langle S^z\rangle$ in a polarized paramagnet can be calculated exactly, $B_{dd}=\frac{\langle S^z\rangle}{\mu_B }\sum_{j\neq i}{\left(2 d_{ij}^{zz}-d_{ij}^{xx}-d_{ij}^{yy}\right)}$. Following Lorentz, we write\cite{Uritskii} ${\bf B}_{dd}={\bf B}'_{dd}+{\bf B}_{Lor}+{\bf B}_{dem}$, where the first term is a sum of dipolar fields within a (Lorentz) sphere with a radius large enough to provide convergence, the second term $B_{Lor}=\frac{1}{3}\chi B_0$ is the Lorentz field that originates from the polarization of the sphere's surface and the third term $B_{dem}=-N\chi B_0$ is the demagnetization field depending on the shape of the sample. By distributing Cu$^{2+}$ spins randomly on both Cu and Zn sites according to the sample stoichiometry, we find $B_{dd}^{'0^\circ}/\chi_{\rm mol}=-1.80\;{\rm T(emu/mol~Cu)^{-1}}$ and $B_{dd}^{'90^\circ}/\chi_{\rm mol}=0.90\;{\rm T(emu/mol~Cu)^{-1}}$. Further, based on the oblate shape of the crystalline grains ($2r/d\sim 10$) we estimate $N^{0^\circ} =0.80(5)$ and $N^{90^\circ} =0.10(3)$,\cite{Akishin} which yields $\left( B_{Lor}+B_{dem}\right)^{0^\circ}/\chi_{\rm mol}=-1.61\;{\rm T(emu/mol~Cu)^{-1}}$ and $\left( B_{Lor}+B_{dem}\right)^{90^\circ}/\chi_{\rm mol}=0.81\;{\rm T(emu/mol~Cu)^{-1}}$.

The effect of the dipolar fields (including demagnetization) is too small to explain the large experimental shifts [see Fig.\ref{figESR}(a) for a comparison of corrected and non-corrected data]. We subtract this contribution and derive ``pure" $g$-shifts [Fig.\ref{figESR}(c)] arising from the symmetric anisotropic exchange. The linear scaling of the shift with macroscopic susceptibility [Fig.\ref{figESR}(d)] can be used to evaluate the magnitude of the exchange anisotropy. Since the Cu-Cl bonds defining the principal axis of the uniaxial local $g$-tensor on each site are tilted from the crystallographic $c$ direction towards the center of the triangle by 43$^\circ$, the direction of the principal axis of $\underline{D}_{ij}$ of each $ij$ bond is not trivial. Therefore, we only inspect the shift data for $\theta=0^\circ$, for which employing the three-fold rotational symmetry of the crystal structure, we derive
\begin{equation}
\Delta g^{0^\circ}(T) = \frac{\chi_{\rm mol}(T)}{N_A g \mu_0 \mu_B^2}4 D'^{cc},
\label{shift2}
\end{equation}
where $D'^{cc}=D_1^{cc}+\frac{1}{2}D_d^{cc}$ corresponds to an effective component of the magnetic anisotropy arising from both dominant bonds $J_1$ and $J_d$.

From the slope $\Delta g^{0^\circ}(T)/\chi_{\rm mol}(T)=-2.0$~(emu/mol Cu)$^{-1}$ we obtain $D'^{cc}=-0.4$~K, yielding the ratio $|D'^{cc}/J_1|=3\%$ with respect to the dominant exchange.  This is somewhat larger than the conventional Moriya estimate $|D/J|\sim (\Delta g/g)^2\sim 2\%$. However, $D'^{cc}$ sums two contributions and, moreover, similarly enhanced values of the symmetric anisotropic exchange interaction were reported before in copper-based systems.\cite{Nidda, Herak} Our results thus suggest that kapellasite is quite close to the isotropic Heisenberg limit. We stress that the DM interaction, which is generally larger than the symmetric exchange anisotropy in copper-based systems (of the order $\Delta g/g\sim 15\%$) could not be assessed in this study, although it may well be present also in kapellasite. Last, we note that the large line shifts in kapellasite observed at low temperatures are due to rather high values of the intrinsic magnetic susceptibility. In contrast, in herbertsmithite and vesignieite, a suppressed susceptibility did not allow an unambiguous assignment of the symmetric exchange anisotropy based on the ESR line shift.
\section{\label{Thermo}Specific heat}
%
%
\begin{figure}
\includegraphics[width=\columnwidth]{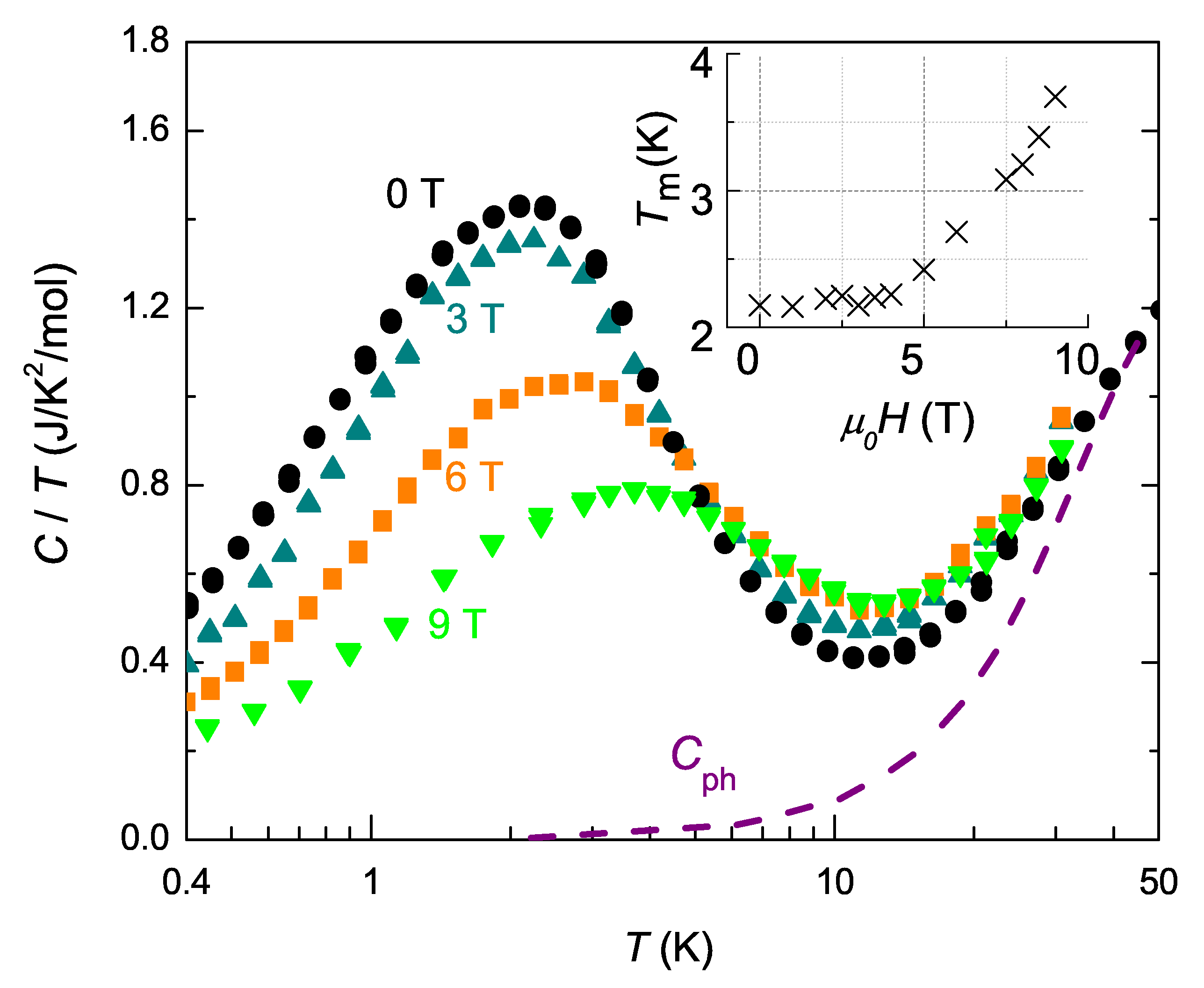}
\caption{\label{CpH} (Color online) Molar specific heat divided by temperature $C/T$ per formula unit vs. temperature and under magnetic field from 0 to 9~T. Below 10~K the phonon contribution $C_{\rm ph}$ is negligible. Inset: The field dependence of the peak position ($T_{\rm m}$) is non-linear, contrary to a Schottky behavior.}
\end{figure}
The specific heat $C(T)$ was measured with a Quantum Design Physical Property Measurement System (PPMS) from 300~K to 0.4~K on thin pressed pellets of kapellasite with a typical mass of 8~mg, in magnetic fields up to 9~T (Fig.\ref{CpH}).
The analysis of $C(T)$ under zero magnetic field has been the subject of two recent publications, where its $T$ dependence was used to refine the model Hamiltonian of kapellasite, based on high temperature series calculations\cite{Kap_letter, Bernu}.
In the following we present its field dependence and confirm that its low-\textit{T} part is magnetic and intrinsic, and thus provides a pertinent quantity for extracting the exchange integrals.

In the absence of a non-magnetic compound isostructural to kapellasite, the phonon contribution is estimated at high temperatures ($50 \leq T \leq 160$~K) with a phenomenological fit made of a linear combination of Debye and Einstein functions.
The lattice contribution is found to be negligible below $T=10~$K (see Fig.\ref{CpH}).
Therefore, the broad peak singled out in $C(T)/T$ around 2~K has clearly a magnetic origin.

The specific heat is found to be insensitive to fields smaller than 3~T, however larger fields cause a shift of the peak towards higher $T$, up to 4~K for $\mu_0 H=9$~T, as well as a decrease of intensity.
This field dependence definitely shows the magnetic character of $C(T)$ at low $T$. The peak that is observed cannot be attributed to the presence of orphan spins which would contribute as a Schottky anomaly in the specific heat as observed for example in herbertsmithite~\cite{Cp_DeVries}. As well as displaying a peak, such a contribution would have a linear dependence in field, contrary to our observations (Fig.\ref{CpH}, inset). This is completely consistent with the fact that the upturn of the susceptibility is not driven by weakly coupled spins but is intrinsic as clearly shown by the perfect match of the measured NMR and ESR shifts with the macroscopic susceptibility.
Therefore, in contrast with other frustrated systems with larger magnetic couplings, the field dependence of $C(T)$ observed here is due to the comparable energy scale between the applied magnetic field and the rather low magnetic exchanges in kapellasite.
As a main consequence, the specific heat for $\mu_0 H=0$~T is characteristic of the model Hamiltonian of kapellasite and is relevant for the evaluation of the exchange interactions as  performed by Bernu \textit{et al.}~\cite{Bernu}.

The $T$ dependence of the magnetic part of $C(T)$ could be reproduced in the range 0.4--10~K within a $J_1 - J_2 - J_d$ model, provided one assumes a $\sim 14$~\% missing entropy, in this $T$-range. In fact, the total magnetic entropy expected for a spin $S=\frac{1}{2}$, $\Delta S=R\ln 2$, is fully recovered only by $T=40$~K, a quite extended regime when compared to the former model. This release of the missing entropy at higher $T$ appears in good agreement with the extended $T$-regime where inelastic neutron scattering studies reveal the presence of magnetic correlations.\cite{Kap_letter}
Two different hypothesis were proposed in Ref.\onlinecite{Bernu} to account for this missing entropy based on the presence of spin vacancies: the existence of weakly coupled spins or the formation of spin singlets. The former scenario can be ruled out, as already discussed, since it would give rise to a linear field dependence of the maximum of the specific heat below 10~K.
The second scenario, namely spin singlet freezing, remains, on the contrary, plausible since the exchange along the diagonal $J_d \sim 15.6$~K  is antiferromagnetic in nature. The identification of the involved spins is still unclear, although the singlet could result from the coupling of a a spin sitting on a kagom\'e site and a spin defect on the center of an hexagon. Their contribution would have to show up at higher temperatures ($T \sim 10$~K), and would therefore remain almost unaffected by fields $\mu_0 H < 9$~T ($\sim 12$~K).
While we are not able to single out the exact behavior of these defects from these measurements, this last scenario is consistent with our experimental results.
It is not clear, however, if a spin singlet can be stabilized in the presence of other large competing interactions.

\section{\label{Dynamics} spin dynamics}
\subsection{1/T$_1$ NMR spin lattice relaxation rate}
\begin{figure*}
\includegraphics[scale=0.5]{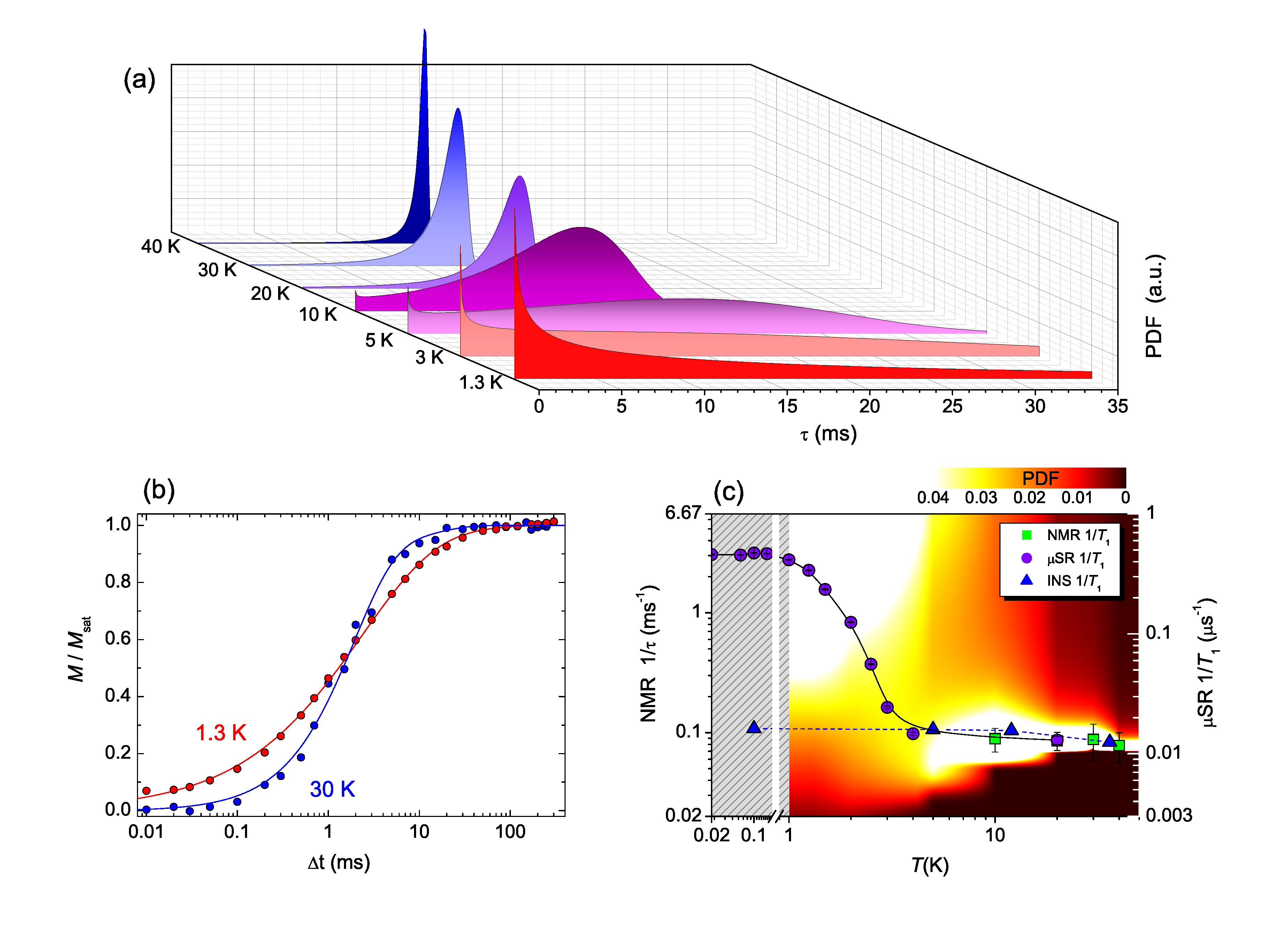}
\caption{\label{NMR_Relax} (a) Probability density functions $\rho(\tau)$ profiles for selected temperatures.
(b) Magnetization-recovery curve at $T = 30$ and $1.3$~K. The conventional relaxation form for $I = 3/2$ is successfully applied at high $T$ (blue line) but non longer valid at 1.3~K where a stretched exponential curve is more appropriate (red line), due to a distribution of relaxation times $\tau$.
(c) Temperature dependence of the spin lattice relaxation rate measured in $\mu$SR ($1/T_1$, right axis, reproduced from \cite{Kap_letter}), in NMR ($1/\tau$, left axis), and calculated from inelastic neutron scattering (INS), see text. The color map represents the intensity of the probability density function $\rho(\tau)$ (left axis) extracted from NMR measurements (grey hatched area is for the absence of NMR data below 1.2~K). Below 10~K, the distribution of $\tau$ is too large to allow a single relaxation time to be defined.}
\end{figure*}
The $^{35}$Cl spin-lattice relaxation rate $1/T_1$ was determined through the saturation recovery method, with a standard $\pi/2 - \Delta t - \pi/2 - \tau - \pi$ pulse sequence ($70 < \tau < 100$ $\mu$s), from 40 to 1.3~K. The measurements were performed at the center of lines (1), (2) and (3), which gave similar average values and distributions (see below) of $T_1$ at low temperature. The evolution of $T_1$ with temperature was then extracted from measurements performed on the line with the highest intensity (1).

Within this $T$-range, the spin dynamics can be divided into two regimes described below: (\textit{i}) a high-$T$ paramagnetic one where $T_1$ remains constant and (\textit{ii}) a low-$T$ fluctuating regime where a broad distribution of relaxation times is observed.

\paragraph*{(i) High $T$ paramagnetic relaxation:}
Above 10 K, the magnetization-recovery curve $M(t)$ is consistent with that expected for a relaxation process of magnetic origin for a nuclear spin $I=\frac{3}{2}$ with a single spin-lattice relaxation time $T_1$\cite{McDowell},
\begin{equation}
M(t,T_1)=M_\mathrm{sat} \left( 1-0.1e^{-t/T_1}-0.9e^{-6t/T_1} \right).
\label{Msaturation}
\end{equation}
Fitting our recovery curves to eq.(\ref{Msaturation}) yields the $T_1$ values plotted in [Fig.\ref{NMR_Relax}(c)] with a constant $T_1 =12(3)$~ms value for $10 < T < 40$~K.

In conventional insulating paramagnets, for $T \gg J$, the spin-lattice relaxation rate $1/T_1$ is indeed temperature independent. The latter was explicitly calculated by Moriya\cite{Moriya} to be
\begin{equation}
1/T_1 = \sqrt{2\pi}\gamma_N^2 g^2 A_\mathrm{hf}^2 S(S+1)/3z_1\omega_e \mbox{, }
\end{equation}
where $z_1 =3$ is the number of Cu$^{2+}$ probed and $\omega_e = J\sqrt{2zS(S+1)/3}/\hbar$ is the exchange frequency with $z=4$ being the number of Cu$^{2+}$ neighbours on the kagom\'e lattice.
Using experimental values  $A_\mathrm{hf} = -4.98$ kOe/$\mu_\mathrm{B}$ for the 3-Cu triangle and the average exchange $J = \theta_{CW} = 9.5$~K, the formula leads to $T_1 = 11$~ms, which is in excellent agreement with the measured $T_1$.
%
\cite{footnote}
\begin{figure}
\includegraphics[width=\columnwidth]{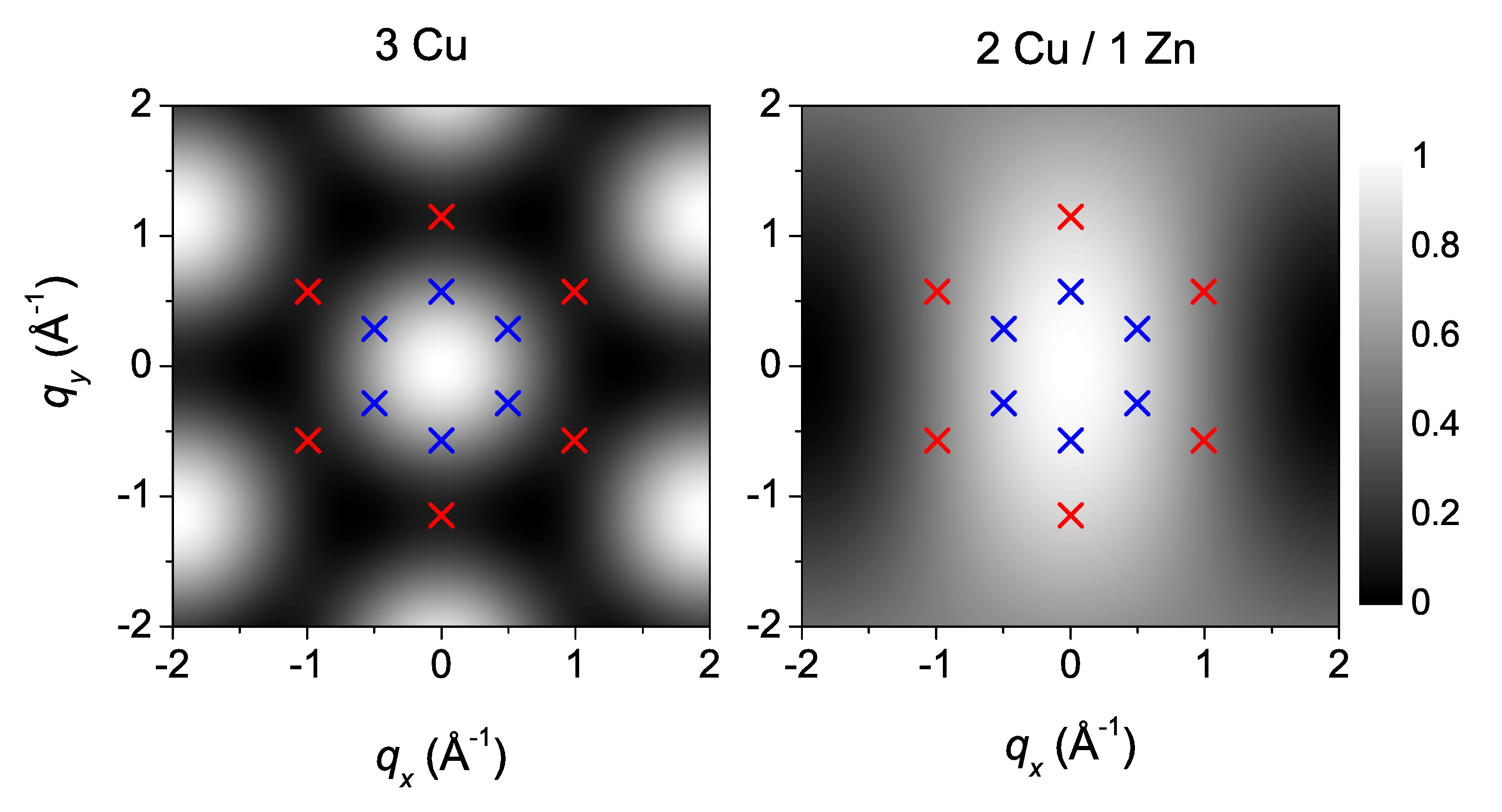}
\caption{\label{formfactor} Intensity of the normalized form factor $|A(q_x, q_y)|^2$ related to site 1 (left) and 2 (right), shown as a grey scale maps. White (black) corresponds to a full (no) transmission. The site 3 does not show any filtering effect. Blue (red) crosses indicate the wave-vectors of the cuboc2 ($q=0$) state.}
\end{figure}

\paragraph*{(ii) Low $T$  relaxation:}
At lower $T$, in the regime where correlations become noticeable, one has to take into account the $q$-dependence of the excitations. Indeed, $T_1$ is given by
\begin{equation}
\frac{1}{T_1} = \frac{\gamma^2}{\mu_B^2}k_B T \sum_q  \left\lvert A(q) \right\lvert^2 \frac{\chi_{\perp}''(q,\omega_{\rm 0})}{\omega_{\rm 0}}
\label{eqT1chi}
\end{equation}
where $A(q)$ acts as a filtering factor for some specific values of $q$, associated here with the triangular geometry. The functions $\left\lvert A(q) \right\lvert^2$ are plotted in Fig.\ref{formfactor} for the various Cl sites. It is notable that at the peak of $S(q)$ detected in neutron experiments~\cite{Kap_letter}, the filtering effect varies by only $\sim 30$~\% from one site to the other, with a minimal transmission of $\sim 60$~\%. In the following we give our results for the Cl site (1) situated at the center of a fully occupied triangle.

If the conventional exponential shape of relaxation is indeed observed above 10~K, the situation is remarkably different at $T=1.3$~K. The saturation recovery curve there had to be fitted using the phenomenological stretched exponential variant of eq.(\ref{Msaturation}) [Fig.\ref{NMR_Relax}(b)],
\begin{equation}
\label{stretched}
\overline{M}(t) = M_\mathrm{sat} \left( 1-0.1e^{-(t/\tau_0)^{\beta}}-0.9e^{-(6t/\tau_0)^{\beta}} \right).
\end{equation}
This points to a broad distribution of relaxation times $\tau$, with a probability density function $\rho(\tau)$, so that
\begin{equation}
\label{eqdistribution}
\overline{M}(t) = \int_0^{+\infty} M(t,\tau)\rho(\tau) d\tau
\end{equation}
Physically, $\tau_0$ is related to the mean of the probability density function while $\beta$ characterizes its dispersion. The probability density function itself can be derived from an inverse Laplace transform of eq.(\ref{eqdistribution}), whose explicit expression can be found elsewhere~\cite{PDF}, and which, following Ref.\onlinecite{Shiroka}, can be numerically evaluated once $\beta$ and $\tau_0$ are known. In order to follow the evolution of the distribution of relaxation times with temperature, we fitted  the magnetization recovery curve with eq.(\ref{stretched}) for each measured temperature. The pair of parameters $(\tau_0, \beta)$ extracted from these fits allows then to compute the $\rho(\tau)$ profiles for $1.3 \leq T \leq 40$~K [Fig.\ref{NMR_Relax}(a)].
Within the investigated $T$-range, $\beta$ is found to change smoothly from 1.0(1) at $T=40$~K to 0.54(4) at $T=1.3$~K.
The results of these fits are presented in the color map of Fig.\ref{NMR_Relax}(c).
Above 10~K the distribution $\rho(\tau)$ is well peaked around $\tau_0$, with $\rho(\tau) \sim \delta(\tau-\tau_0)$,
which legitimately defines a single relaxation time $T_1 = \tau_0 \sim 12$~ms, with $\beta \rightarrow 1$.
It then asymmetrically broadens at lower $T$, exhibiting a $\tau \rightarrow 0$ divergence when $T \rightarrow 0$, while $\beta$ approaches smoothly its lowest value of $\sim 0.5$.
The same qualitative behavior was observed for all Cl sites.

\subsection{LF $\mu$SR relaxation}
\begin{figure}
\includegraphics[width=\columnwidth]{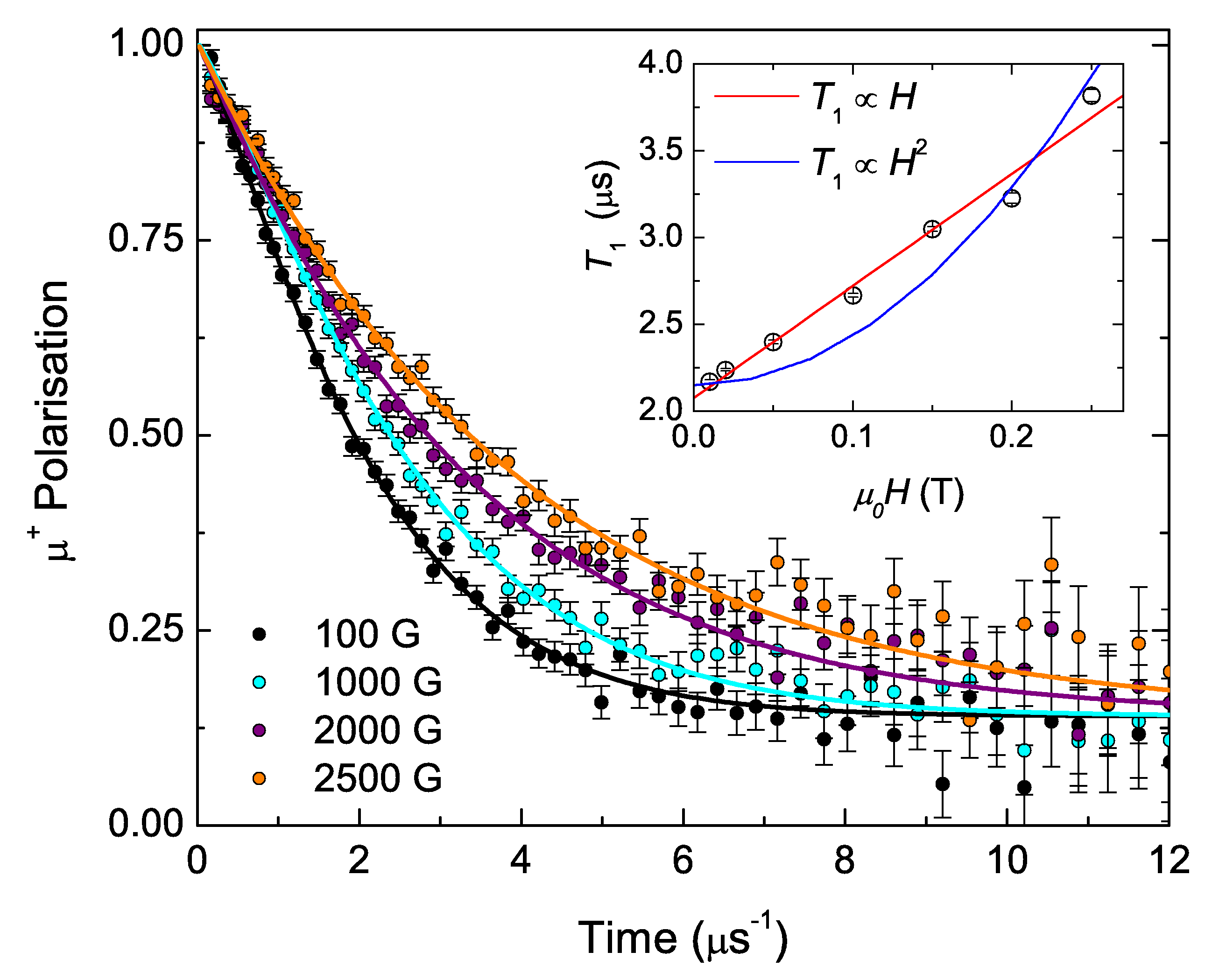}
\caption{\label{T1musr} (Color online) Muon polarization under longitudinal applied magnetic fields $\mu_0 H = 0 -2500$~G, at $T=60$~mK. Inset: $H$ dependence of the spin-lattice relaxation time $T_1$.}
\end{figure}
Complementary information on the spin dynamics can be obtained from $\mu$SR measurements in longitudinal field, since it directly probes the spectral fluctuation density. As discussed in Ref.\onlinecite{Kap_letter}, the $\mu^+$ polarisation is described by $P(t) = fe^{-t/T_1} +(1-f)$, where $1-f = 0.14$ is a slowly relaxing muons fraction completely decoupled for $\mu_0 H > 100$~G and thus responsible for the non-zero value observed at long time~\cite{Kap_letter} (Fig.\ref{T1musr}).
The application of a small longitudinal field $\mu_0 H = 100$~G also enables one to negate the static nuclear contribution and allows the electronic fluctuations to be probed separately.

Similarly to NMR, a constant relaxation rate $1/T_1^{\mu}$ is observed for $T > 5$~K [Fig.\ref{NMR_Relax}(c)].
Assuming local fluctuations, the spin-lattice relaxation times are linked through the relation $T_1 / T_1^{\mu} = ( \gamma_{\mu}A_{\mu} / \gamma_{n} A_\mathrm{hf} )^2$.
At $T = 20$~K, the experimental scaling factor is found to be $T_1 / T_1^{\mu} \sim 150$, leading to the muon coupling constant $A_{\mu} = 1.88$~kOe/$\mu_B$, close to the dipolar coupling estimated to be $A_{\mu}^{\rm dip} \sim 1.3$~kOe/$\mu_B$ at the principal OH$^-$ muon site. Thus, the spin fluctuations probed in $\mu$SR above 10~K also originate from a paramagnetic regime.

A good agreement is observed for the 3~K onset-$T$ where the relaxation rate departs from the paramagnetic limit while the evolution at lower $T$ follows qualitatively that of the probability density function profiles extracted from $^{35}$Cl NMR, as underlined by Fig.\ref{NMR_Relax}(c).

However, contrary to NMR, a broad distribution of relaxation times does not appear clearly in $\mu$SR experiments. Should the same distribution be observed, the $\mu^+$ polarisation for $\mu_0 H = 100$~G  would vary as $P(t) = e^{-(t/T_1)^\beta}$ with $\beta \sim 0.5$ at low $T$. Allowing a stretched exponential relaxation does not appreciably improve the quality of the fit. The $\beta$ exponent is ill-defined but its value is found to remain above 0.7 in the whole $T$-range below 4~K which would suggest a moderate distribution of relaxation times as probed by the muon.

Lastly, we comment on the field-dependence of the relaxation rate. In disordered spin systems with a single relaxation time and exponentially decaying correlations, the spin-lattice relaxation time $T_1$ exhibits a field square dependence $T_1 \propto H^2$, characteristic of a Lorentzian spectral density.\cite{Baker}
Such a spectral-weight function, which is  centered at zero energy, is commonly used to describe conventional fluctuations in the paramagnetic regime.
Here, a power law dependence $T_1 \propto H^\mathrm{\alpha}$ is observed at 60~mK, with $\alpha \simeq 1$ (Fig.\ref{T1musr}, inset).
This observation is inconsistent with the existence of a single time scale, as underlined by our NMR measurements, and suggests a more exotic spectral density, such as the one at play in the spin-liquid phase of the $S=\frac{1}{2}$ antiferromagnetic chain.\cite{Pratt2006}

The contrast between the low-$T$ relaxation extracted from the two NMR and $\mu$SR techniques is discussed in the following section.

\section{Discussion}
\subsection{\label{disorder}Effects of disorder}
%
%
In this section, we discuss the information that can be extracted from the NMR spectra and the related impact of defects, that is deviations from the ideal kagom\'e lattice, on the ground-state of kapellasite.
Let us first recall that the structural refinements yield the chemical formula \kapp ~which points to both the dilution of the kagom\'e lattice (27~\%) and a substantial occupation of the Zn site by Cu (12~\%).

\begin{figure}
\includegraphics[width=0.9\columnwidth]{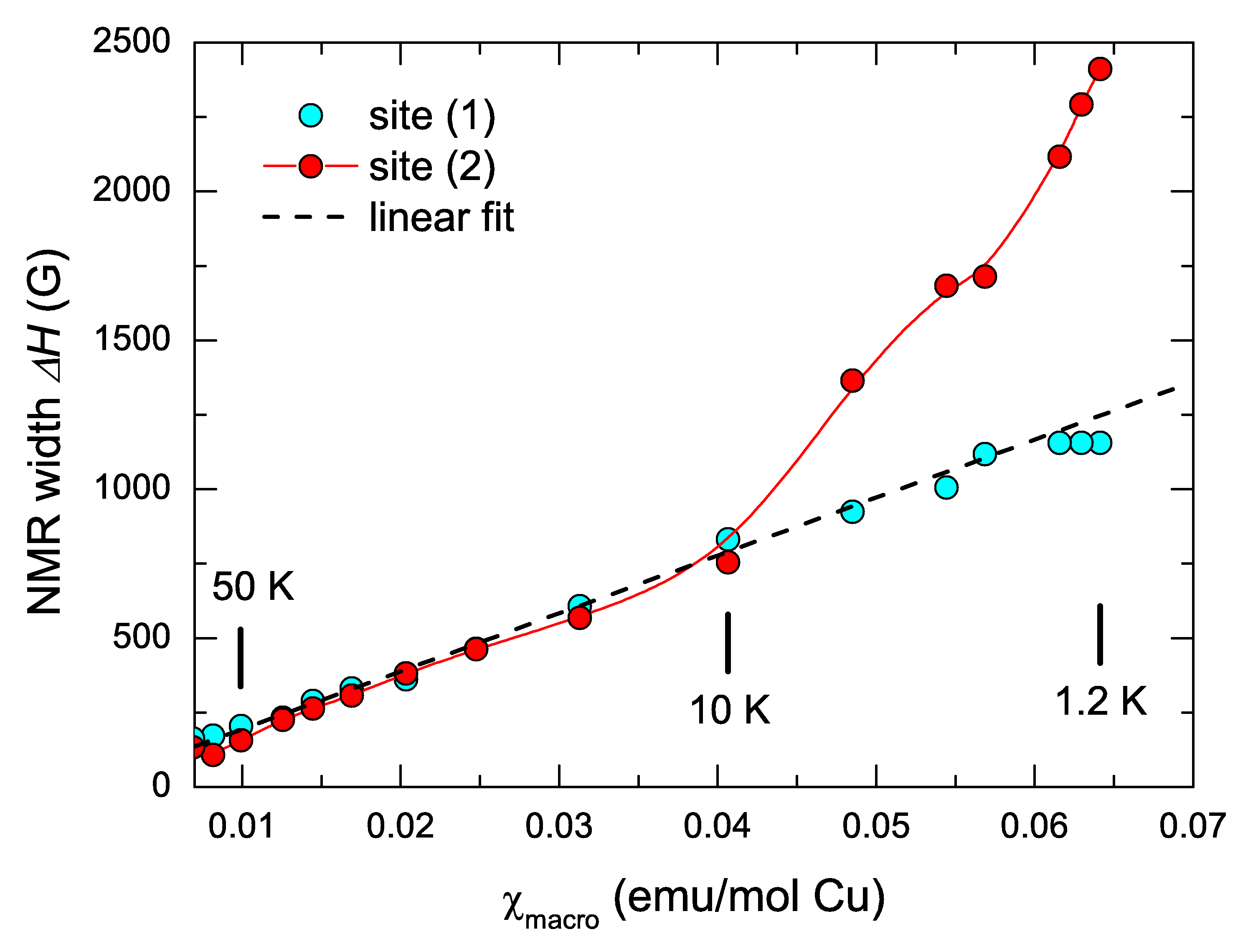}
\caption{\label{NMRwidth} (Color online) FWHM of the NMR lines corresponding to site(1) and (2), from the spectra displayed on Fig.\ref{NMRchi}. Temperature is an implicit parameter.}
\end{figure}

Spin vacancies are rather common in experimental materials, e.g. in the kagom\'e antiferromagnet where, for a nn-antiferromagnetic Hamiltonian, they are known to create a spin singlet on the sites first neighbours of the impurity and induce a staggered susceptibility at larger distances.\cite{Dommange2003,Rouso2009,Poilblanc2010} In the case of kapellasite, where the scheme of interactions is different, little is known about such an impact of the spin vacancy. From a general view point, magnetic defects are expected to generate a distribution of susceptibilities. In addition, the rapid crystallisation of kapellasite during the synthesis process\cite{Kap_formula} may affect the homogeneity of the defect distribution, and could further influence the disorder.
The latter can yield a distribution of hyperfine constants. Although we expect the distribution of susceptibilities to dominate  the linewidth of our spectra, $\Delta H \sim \Delta (A_{\rm hf}\chi)$, especially in the low $T$ range, attributing all the width to a structural disorder-induced distribution of hyperfine constants would lead to a scaling $\Delta H \sim \chi$.
An upper bound of $\Delta A_{\rm hf}/A_{\rm hf} \sim 33$~\% can then be extracted from the linear slope of the plot of $\Delta H$ versus $\chi$ in Fig.\ref{NMRwidth}. In the low-$T$ range, we note that, for site (2), the linear scaling is severely disrupted below 10~K which rather argues in favour of a distribution of susceptibilities, as expected.

In the context of a large dilution, one could wonder whether the absence of any spin freezing really reflects the spin liquid physics of the ideal $J_1-J_2-J_d$ Heisenberg Hamiltonian. We first recall that the level of dilution of $d=27$~\% is still above the percolation threshold of the kagom\'e lattice $p_c \simeq 0.652$,\cite{Percolation1,Percolation2} i.e. that $d < d_c = 1-p_c \simeq 0.35$. On the contrary, delafossite LaCuO$_{2.66}$, with a 33~\% ($\lesssim d_c$) diluted kagom\'e lattice, displays an inhomogeneous spin frozen ground-state.\cite{LaCuO2.66} While the extreme two-dimensional character of kapellasite could also be invoked to support the absence of freezing at finite temperature, the Mg-analogue of kapellasite, haydeite, has a clear transition at $T \sim 4$~K.\cite{Kap_formula,Hayd_Chu} This gives support to the intrinsic character of the spin liquid behavior which is found in kapellasite.

To date, the existence of such a spin liquid phase is supported within two theoretical approaches: ($i$) a Schwinger-Boson mean-field model by Messio \textit{et al.},  which reproduces most of the experimental observations,\cite{Kap_letter} including the evolution of the $\mu$SR relaxation rate with temperature, but inherently fails to reproduce a gapless behavior as observed both in neutron and our experiments; and ($ii$) a pseudofermion functional renormalization group approach by R. Suttner \textit{et al.}.\cite{J1Jd_theory} There, kapellasite would locate in a small non-frozen region between the cuboc and the ferromagnetic phase, in a slightly different $J_1-J_2$ Heisenberg model.

\subsection{Inhomogeneous dynamics}
%
We now turn to the spin dynamics in the low temperature phase and discuss the intriguing difference observed between the relaxation probed by NMR and $\mu$SR.
The form factor $\left\lvert A(q) \right\lvert^2$ in eq.(\ref{eqT1chi}) could be a potential source of disagreement between the two measured relaxations. The muon most likely sits in a non-symmetric site and is dominantly coupled to Cu$^{2+}$ via dipolar interaction while the $^{35}$Cl nucleus, located at a symmetric site above a Cu$^{2+}$ triangle probes each Cu at the corners of this triangle through hyperfine couplings.

While the distribution thought to be of hyperfine couplings mentioned previously could be responsible for the distribution of $T_1$ seen only in NMR, this can be ruled out as the broadening of the distribution function is clearly temperature dependent [Fig.\ref{NMR_Relax}(a)]; a distribution of hyperfine constants remains $T$-independent.

The $q$-dependence of the form factor may also filter out some modes in NMR as compared to $\mu$SR. Let us recall that the different sites are clearly resolved in the NMR spectra, and that the same stretching of the relaxation is observed at each site at low temperature. Since the intensity of the form factors differs from one site to another (see Fig.\ref{formfactor}), it is unlikely that the $q$-dependence of $\left\lvert A(q) \right\lvert^2$ is responsible for the $T_1$ distribution.

We are thus left with the firm conclusion that the distribution of $T_1$ observed in NMR mainly comes from the fluctuation spectrum.
The difference between NMR and $\mu$SR relaxations could be ascribed to the difference of these couplings rather than being purely explained by the different time-scale probed with each technique. Indeed, for both techniques, $T_1^{-1}$ scales with $A^2$ -- $A$ is the coupling constant -- so that the long time tail of the $T_1$ distribution detected through NMR would yield unobservable relaxation times in $\mu$SR. Cutting off this tail by restricting the integration range of the probability density function extracted from NMR over $[0, \alpha]$ with $\alpha < +\infty$,  yields a distribution of $1/T_{1}^{\mu} = \lambda$ [see eq.(\ref{eqdistribution})] still too large to account for the smaller breadth of relaxation times ($\beta \geq 0.7$) observed in $\mu$SR. It is therefore necessary to invoke very local modes that give distinct relaxations characteristics to the various sites probed by NMR and which have a distribution that is smoothed out through the long range character of the dipolar coupling between the muon probe and the fluctuating Cu$^{2+}$ spins.

One of the characteristic features of the low $T$ phase of kapellasite is the well localized intensity present in the inelastic channel measured by inelastic neutron scattering (INS) at the $M$ point of the Brillouin zone. Both INS and NMR relaxation probe the dynamical susceptibility $\chi''(q,\omega)$, yet NMR has no $q$-resolution (except via the form factor) and probes the $\omega \rightarrow 0$ region.
In order to extract some information from their comparison, we convert $\chi''(q,\omega)$ measured by INS in an equivalent $1/T_1$ by integrating $\chi''$ in the ranges $q=[0.2,0.8]$~{\AA} and $\omega=[0.4,0.8]$~meV. The values obtained are scaled so as to match the high $T$ NMR and $\mu$SR ones. The result is plotted on Fig.\ref{NMR_Relax}(c).
We note that no dramatic filtering effect is expected when applying the NMR form factor $\left\lvert A(q) \right\lvert^2$ of different sites, in agreement with the non-zero value of the form factors at the cuboc2 $q$-positions (Fig.\ref{formfactor}).
The $T$ evolution of $T_1$ generated from the INS data is in good agreement with the flat dependence observed by local probes above 10~K, but fails to reproduce any distribution at lower $T$. Therefore, we can conclude that at low $T$ the NMR relaxation is dominated by extra low-energy modes that are inaccessible to INS. The low energy fluctuation spectrum would then have spectral weight relative not only to the cuboc2 mode but also at many different $q$ positions, not filtered out by the form factors.

It is interesting to note that the cuboc2 phase is chiral and consequently could allow for the observation of a chiral phase transition at finite temperature, studied theoretically in details in Refs.\onlinecite{Messio_chiral,Domenge2008}. This transition is combined with a proliferation of topological point defects ($\mathbb{Z}_{\rm 2}$ vortices), which preferentially nucleate around the walls separating domains with opposite chirality. Such excitations were previously suspected to be at the origin of the unconventional spin dynamics observed in the Heisenberg triangular antiferromagnet NaCrO$_2$\cite{NaCrO2_Olariu}.
Further work would be required to confirm that the peculiar low-energy excitations observed in kapellasite are indeed a signature of such an exotic transition.

\section{Conclusion}
We reported the experimental investigation of the $S=1/2$ Heisenberg kagom\'e system kapellasite, where the two main competing interactions $J_1$ and $J_d$ conspire to stabilize a spin liquid state at low $T$ with spin correlations of the cuboc2 type.
$^{35}$Cl NMR data have confirmed the previously established level of dilution of 27\% of the kagom\'e lattice and further evidenced its random character. Surprisingly, an identical local magnetic susceptibility is probed at high temperature for each Cu$^{2+}$ spin, irrespective of its neighbourhood.
This is a strong evidence of the homogeneity of the physics in the system, despite the substantial level of disorder.
Additional improvements in our understanding of the disorder effect would require a set of samples with various dilution, which remains at the moment a real chemical challenge, due to the metastability of the kapellasite phase.
ESR showed that the Heisenberg model is relevant to kapellasite and established a moderate magnetic anisotropy of the symmetric anisotropic exchange type of $|D'^{cc}/J_1|=3\%$ while no conclusion can be given about the Dzialoshinskii-Moriya anisotropy.
The intrinsic character of the specific heat and of the magnetic susceptibility are thoroughly assessed which gives a solid experimental credit to the theoretical procedure applied by Bernu \textit{et al.}\cite{Bernu} to extract the exchange couplings of kapellasite from high temperature series fits.
Unconventional spin dynamics is consistently revealed by NMR, $\mu$SR and inelastic neutron scattering\cite{Kap_letter} upon cooling below $T=5$~K, where spin correlations effectively built up. The concomitant distribution of relaxation observed in NMR is then due to the emergence of extra low energy modes, inaccessible to the neutron time scale. It could either be a complex signature of the effect of disorder, or of the specific excitation spectrum of the low-$T$ cuboc2 chiral phase.

In conclusion, while a ferromagnetic kagom\'e lattice is not frustrated, the existence of a further nearest-neighbour exchange along the diagonal $J_d$ in kapellasite, comparable to $|J_1|$, restores magnetic frustration via the competition of interactions. The effect of frustration through the geometry of interactions was only relatively recently explored on the kagom\'e lattice \cite{J1Jd_theory}. Up to now, most of the work on kagom\'e magnets has been dedicated to the pure antiferromagnet Heisenberg model and the consideration of the effects of additional energy scales is still in its infancy. Second neighbour interactions have also been argued to be necessary to explain the neutron data on herbertsmithite single crystals.\cite{J1Jd_theory}  Kapellasite gives an order of magnitude for such interactions and more broadly opens the way towards the experimental study of the phase diagram of the $J_1 - J_2 - J_d$ Heisenberg model which displays non-coplanar chiral ordered phases and non-magnetic regions.
The discovery of its spin liquid state certainly deserves further attention in order to reach a more comprehensive understanding, and provide new exciting perspectives to explore quantum spin liquids physics on the kagom\'e lattice.

\subsection*{Acknowledgements}
EK acknowledges very useful discussions with B. Bernu. This work was supported in part by the European
Commission under the 6th Framework Programme Contract No.\ RII3-CT-2003-505925, by the french Agence Nationale de la Recherche under grants ANR-09-JCJC-0093-01, ANR-SPINLIQ-86998, ANR-HFM-86998, by the PHC Proteus program No. 24322SF and by Universit\'e Paris-Sud, grant MRM PMP.  A.Z. acknowledges the financial support of the Slovenian Research Agency, projects No. BI-US/14-45-039. The NHMFL is supported by NSF Cooperative Agreement No. DMR-1157490 and by the state government of Florida.

\appendix
\section{\label{NMRappendix} $^{35}$Cl NMR measurements}
\subsection{Oriented sample and experiments}
NMR experiments were performed on an oriented powder of kapellasite, aligned along the $c$-axis. This sample was prepared by mixing together 200 mg of kapellasite powder, prepared following Ref.\onlinecite{Colman2008}, with a stycast epoxy. The mixture was then placed in a cylindrical teflon mold and cured under a magnetic field of 7~T for 15~h.
The level of orientation was estimated to be $\sim90$\% from the comparison of intensities between the non-oriented and oriented NMR spectrum.
Therefore, the oriented sample behaves like a single crystal in the $c$-axis direction, i.e. perpendicular to the kagom\'e planes. This orientation allows the observation of narrow NMR lines instead of conventional complex powder lineshape due to  quadrupolar effects.

NMR experiments were performed on the most abundant Cl isotope, $^{35}$Cl (abundance of 75.8~\%), with a nuclear spin $I=3/2$ and a gyromagnetic ratio $\gamma_{n} = 4.1716$~MHz/T.
There is only one Cl crystallographic site within the structure, which is coupled to three Cu$^{2+}$ ions of one triangle belonging to the kagom\'e plane [Fig.\ref{structure}(d)].
NMR measurements were made both in fixed and sweep field configurations, under various magnetic fields from 3 to 13~T. A $^4$He flow cryostat was used to reach temperatures from 1.2 to 300~K. 
\subsection{Nuclear Hamiltonian}
$^{35}$Cl is a quadrupolar nucleus ($I=3/2$), hence one has to consider two sets of interactions resulting from the coupling of the nucleus to its magnetic and charge environments. The nuclear Hamiltonian can be written\cite{slichter,cohen}
\begin{equation}
\mathcal{H} = \mathcal{H}_\mathrm{mag} + \mathcal{H}_\mathrm{quad} \\
\end{equation}
with
\begin{align}
\label{Hm}  \mathcal{H}_\mathrm{mag} &= -\gamma_n \hbar \mathbf{H_0} \cdot \mathbf{I} -\gamma_n \hbar \mathbf{H_0}  \cdot \widetilde{K} \cdot \mathbf{I} \\
\label{Hq}  \mathcal{H}_\mathrm{quad} &= \frac{e^2qQ}{4I(2I-1)}\left[(3I_{z'}^2-I^2) + \eta (I_{x'}^2 - I_{y'}^2)\right]
\end{align}
The first term of eq.(\ref{Hm}) corresponds to the nuclear Zeeman effect under a static applied field $\mathbf{H_0}$. The second term accounts for the magnetic interaction of the nuclear spin with its local environment through the shift tensor $\widetilde{K}$. Its $T$-dependent part directly probes the spin-susceptibility of the surrounding magnetic moments.

The quadrupolar Hamiltonian [eq.(\ref{Hq})] is  characterized by two parameters: $\nu_{q}=3e^2qQ/2I(2I-1)h$, which results from the coupling of the nuclear quadrupole moment $Q$ with the electrostatic field gradient (EFG)  characterized by its $q$ value along its principal $z'$ axis, and its asymmetry, $\eta$, in the $(x',y')$ plane ($0\leq \eta \leq 1$).

For an ``ideal" structure of kapellasite, {\it i.e.} with perfect stoichiometry and no local distortion which might result from the Zn/Cu substitution, a threefold axial symmetry exists at the Cl site. This simplifies considerably both the hyperfine tensor and the quadrupolar Hamiltonian which then have the same principal axis along $z$ and a rotational invariance in the ($x$, $y$) plane, hence $K_x=K_y$ and $\eta = 0$.

Under the applied fields of a few teslas,  $\mathcal{H}_\mathrm{quad}$ is a small perturbation as compared to  $\mathcal{H}_\mathrm{mag}$ for the experimental case considered in this work. In first order, the $(2I+1)$ energy levels $E_m$ write $E_m = \langle m |\mathcal{H} | m\rangle$, where $-I \leq m \leq I$ and $| m\rangle$ is the eigenstate of the Zeeman Hamiltonian for an applied field $\mathbf{H_0}$.
This energy splitting gives rise to the $2I$ resonance frequencies $\nu_{m-1,m} = (E_{m-1} - E_m)/h$ detected in NMR.
Taking into account the previous symmetry arguments and expressing $\mathbf{H_0}$ in spherical coordinates, $\mathbf{H_0} = H_0(\sin\theta \cos\phi, \sin\theta \sin\phi, \cos\theta)_{x,y,z}$, we obtain for the ``ideal" case:
%
\begin{align}
\label{nu}
\nu_{m-1,m} =          &~~~\nu_0 \\
\nonumber        & +\nu_0 (K_{x,y}\sin^2\theta + K_z\cos^2\theta) \\
\nonumber        & + \left(m-\frac{1}{2}\right)\frac{\nu_{q}}{2} \left[3\cos^2\theta -1 \right]
\end{align}
%
where $\nu_0$ is the resonance frequency of the bare nucleus $\gamma_n H_0$ corrected by the usual $T$-independent chemical shift. For $^{35}$Cl-NMR (I = 3/2), one obtains three resonances: one with no quadrupolar shift to the first order approximation and is only governed by magnetic effects -- the central line at $\nu_{-1/2,1/2}$~--, and two satellites $\nu_{1/2,3/2}$ and $\nu_{-3/2,-1/2}$. With the applied field along $z$ axis ($\theta=0$\degree) and in the case of axial symmetry ($\eta = 0$), we simply derive:
\begin{align}
\label{nu1} \nu_{-1/2,1/2}            &=      \nu_0 (1 + K_z) \\
\label{nu2} \nu_{1/2,3/2}            &=      \nu_0 (1 + K_z) + \nu_q\\
\label{nu3} \nu_{-3/2,-1/2}          &=       \nu_0 (1 + K_z) - \nu_q
\end{align}
Similar formula can be obtained for the applied field perpendicular to the $z$ axis.
\begin{align}
\label{nu1} \nu_{-1/2,1/2}            &=      \nu_0 (1 + K_{x,y}) \\
\label{nu2} \nu_{1/2,3/2}            &=      \nu_0 (1 + K_{x,y}) - \nu_q /2\\
\label{nu3} \nu_{-3/2,-1/2}          &=       \nu_0 (1 + K_{x,y}) + \nu_q /2
\end{align}
Assuming as a first approximation that the susceptibility is isotropic, we can directly extract the variation of the shift when the field is applied along $z$ direction, $K_z$, from the variation of the susceptibility through $\delta K_z(T) = A_{\mathrm{hf},z} \cdot \delta\chi_\mathrm{spin}(T)$ where $A_{\mathrm{hf},z}$ is the hyperfine coupling for $\mathbf{H_0}\parallel z$.

As a final note to this section, we underline that any local structural distortion will introduce deviations from these simple formula ($\eta \neq 0$ and different principal axis for the EFG and shift tensor not aligned along $z$).


\begin{thebibliography}{99}
\bibitem{footnote}
It should be noted that, since $1/T_1 \sim A_\mathrm{hf}^2/z_1\sqrt{z} \sim z_1/\sqrt{z}$, different $T_1$ values are expected for the pure and diluted Cl sites due to the dependence through $ \sim z_1/\sqrt{z}$.
Calculations using the statistical average parameter $\overline{z}$ yield $T_1$ varying from 10 to 15 ms, and thus differing by $\sim 25$\% from site (1). In this high-$T$ regime, our accuracy on the $T_1$ value is not high enough to resolve such a difference.
\end{thebibliography}
\end{document}